\documentclass[lettersize,journal]{IEEEtran}
\usepackage{amsmath,amsfonts}
\usepackage{algorithmic}
\usepackage{algorithm}
\usepackage{array}
\usepackage{textcomp}
\usepackage{stfloats}
\usepackage{url}
\usepackage{verbatim}
\usepackage{graphicx}
\usepackage{cite}
\usepackage{multirow}
\ifCLASSOPTIONcompsoc
    \usepackage[caption=false, font=normalsize, labelfont=sf, textfont=sf]{subfig}
\else
\usepackage[caption=false, font=footnotesize]{subfig}
\fi
\begin{document}

\title{AVSim - Realistic Simulation Framework for Airborne and Vector-Borne Disease Dynamics}


\author{
        Pandula Thennakoon,
        Mario De Silva$^{\star}$,
        Mahesha Viduranga,
        Sashini Liyanage,
        Roshan Godaliyadda, \\
        Mervyn Parakrama Ekanayake,
        Vijitha Herath,
        Anuruddhika Rathnayake,
        Ganga Thilakarathne, \\
        Janaka Ekanayake,~\IEEEmembership{Senior Member,~IEEE,}
        Samath Dharmarathne
        \vspace{0.1in} \\
        \textit{This work has been submitted to the IEEE for possible publication. Copyright may be transferred without notice, after which this version may no longer be accessible.}

\thanks{$\star$ Corresponding author: Mario De Silva(e19463@eng.pdn.ac.lk).}

\thanks{Pandula Thennakoon, Mario De Silva, Mahesha Viduranga, Roshan Godaliyadda, Mervyn Parakrama Ekanayake, Vijitha Herath, Janaka Ekanayake is with the Department of Electrical and Electronic Engineering, University of Peradeniya, Sri Lanka}

\thanks{Mahesha Viduranga, Sashini Liyanage, Roshan Godaliyadda, Mervyn Parakrama Ekanayake, Vijitha Herath, Janaka Ekanayake is with the Multidisciplinary AI Research Center, University of Peradeniya, Sri Lanka}

\thanks{Sashini Liyanage is with the Department of Computer Engineering, University of Peradeniya, Sri Lanka}

\thanks{Anuruddhika Rathnayake, Samath Dharmarathne is with the Faculty of Medicine, University of Peradeniya, Peradeniya, Sri Lanka}

\thanks{Ganga Thilakarathne is with the Institute of Policy Studies, Colombo, Sri Lanka}
}

\markboth{IEEE TRANSACTIONS ON SYSTEMS, MAN, AND CYBERNETICS: SYSTEMS, 2025}%
{Shell \MakeLowercase{\textit{et al.}}: A Sample Article Using IEEEtran.cls for IEEE Journals}


\maketitle

\begin{abstract}
Computational disease modeling plays a crucial role in understanding and controlling the transmission of infectious diseases. While agent-based models (ABMs) provide detailed insights into individual dynamics, accurately replicating human motion remains challenging due to its complex, multi factorial nature. Most existing frameworks fail to model realistic human motion, leading to oversimplified and less realistic behavior modeling. Furthermore, many current models rely on synthetic assumptions and fail to account for realistic environmental structures, transportation systems, and behavioral heterogeneity across occupation groups. To address these limitations, we introduce AVSim, an agent-based simulation framework designed to model airborne and vector-borne disease dynamics under realistic conditions.
A distinguishing feature of AVSim is its ability to accurately model the dual nature of human mobility (both the destinations individuals visit and the duration of their stay) by utilizing GPS traces from real-world participants, characterized by occupation. This enables a significantly more granular and realistic representation of human movement compared to existing approaches.
Furthermore, spectral clustering combined with graph-theoretic analysis is used to uncover latent behavioral patterns within occupations, enabling fine-grained modeling of agent behavior.
We validate the synthetic human mobility patterns against ground-truth GPS data and demonstrate AVSim’s capabilities via simulations of COVID-19 and dengue. The results highlight AVSim’s capacity to trace infection pathways, identify high-risk zones, and evaluate interventions such as vaccination, quarantine, and vector control with occupational and geographic specificity.
\end{abstract}

\begin{IEEEkeywords}
Agent-based models, social systems, epidemiology, unsupervised clustering, COVID-19, Dengue
\end{IEEEkeywords}

\section{Introduction}
\IEEEPARstart{D}{isease} modeling is crucial in modern society for predicting and managing infectious disease spread, enabling proactive public health policies. It optimizes resource allocation, informs targeted interventions, and enhances forethought for emerging epidemic threats. The need for robust computational frameworks to model epidemic dynamics is highlighted by the rapid spread of airborne and vector-borne diseases. Starting from December 2019, the COVID-19 pandemic lasted over three years, resulting in more than 700 million cases and claiming over 7 million lives \cite{whoCOVID19Deaths}. This crisis highlighted the critical importance of timely interventions and coordinated efforts by individuals, healthcare systems, and governments to control disease outbreaks and prevent them from escalating into pandemics. Beyond COVID-19, vector-borne diseases, transmitted by living organisms that carry infectious pathogens between humans or from animals to humans, pose a significant public health challenge, causing over 700,000 deaths annually. For instance, in 2024, the World Health Organization (WHO) estimated approximately 249 million global cases of malaria and 100–400 million cases of dengue each year \cite{whoVectorborneDiseases}. Many vector-borne and airborne diseases, including those similar to COVID-19, are preventable through protective measures and effective community mobilization. Modeling these disease dynamics helps to identify hot spots, optimize interventions and resource allocation, and further enable policymakers to implement proactive, data-driven strategies.

Effective intervention and prevention strategies in epidemiology require modeling disease spread under realistic conditions. Human mobility, behavioral patterns, and public modes of transportation significantly influence the speed and extent of pathogen transmission \cite{Weligampola2023}. Different behaviors and transport modes affect disease spread in varied ways. However, many models either overlook these variations \cite{Trivedi2021, Fan2024} or rely on oversimplified assumptions, such as treating transportation as random interactions aggregated with activities like shopping or social events \cite{COVASIM, OPENABM}, failing to reflect real-world complexities. Furthermore, previous methods fail to capture realistic human motion patterns and underlying social dynamics. It is well known that modeling human behavior is a complex task due to the inherent complexity and unpredictability of human systems. Humans exhibit diverse behaviors influenced by factors like profession, age, culture, environment, and personal preferences. Not only that, but social dynamics arise from complex, nonlinear interactions among individuals and environments. Therefore, it is challenging to model this type of dynamical system and predict outcomes without a detailed understanding of the feedback loops and control mechanisms within the system.

\begin{table*}[h]
    \centering
    \caption{Comparison of disease simulators found in the literature with ours.}
    \label{tab:disease_simulator_comparison}
    \resizebox{\textwidth}{!}{%
    \begin{tabular}{|l|l|c|c|c|c|c|c|c|c|c|c|c|c|}
    \hline
    \textbf{Simulation Method} & \textbf{Simulator} & \rotatebox{90}{\textbf{Realistic Human Motion}} & \rotatebox{90}{\textbf{Real Mobility Data Usage}} & \rotatebox{90}{\textbf{Realistic Environment}} & \rotatebox{90}{\textbf{Transport System}} & \rotatebox{90}{\textbf{Testing Protocol}} & \rotatebox{90}{\textbf{Vaccination Strategy}} & \rotatebox{90}{\textbf{Containment Strategy}} & \rotatebox{90}{\textbf{Hygiene Protocol}} & \rotatebox{90}{\textbf{Disease Spread Tracing} \hspace{0.3em}} & \rotatebox{90}{\textbf{User-Friendly Interface}} & \rotatebox{90}{\textbf{Detailed Agent Modeling}} & \rotatebox{90}{\textbf{Climatic Conditions}} \\
    \hline
    \multirow{6}{*}{\textbf{Vector-borne Diseases}} & Network-patch methodology for adapting disease models \cite{patch-2014} & \checkmark & & & & & & & & & & \checkmark & \checkmark\\ \cline{2-14}
     & Agent-Based Simulation of Dengue Fever Spread \cite{dengueMohomad} 
 & & & \checkmark & & \checkmark & & & & & \checkmark & \checkmark & \\ \cline{2-14}
     & Climate-based dengue model in Semarang, Indonesia \cite{climate_baseed_dengue_model} & & & \checkmark & & \checkmark & & & & & & & \checkmark \\ \cline{2-14}
     & ABM for vaccine efficacy's influence on public health \cite{Perkins082396} & \checkmark & & \checkmark & & \checkmark & \checkmark & & \checkmark & & & \checkmark & \\ \cline{2-14}
     & ABM driven by rainfall to study chikungunya outbreak \cite{DOMMAR201461} & \checkmark & & \checkmark & & \checkmark & & & & \checkmark & \checkmark  & \checkmark & \checkmark  \\ \cline{2-14}
     & \textbf{AVSim - Vector-borne model (Our Approach)} & \checkmark & \checkmark & \checkmark & & \checkmark & & \checkmark & \checkmark & \checkmark & \checkmark & \checkmark & \checkmark \\ \hline
    \multirow{10}{*}{\textbf{Air-borne Diseases}}
    
     & Covasim \cite{COVASIM} & &  & & & \checkmark & \checkmark & \checkmark & \checkmark& \checkmark & \checkmark & &  \\ \cline{2-14}
    
     & OpenABM-COVID-19 \cite{OPENABM} &  &  & & & \checkmark & \checkmark & \checkmark & \checkmark & \checkmark & & &   \\ \cline{2-14}

     & PDSIM \cite{Weligampola2023} & \checkmark &  & & \checkmark & \checkmark & \checkmark & \checkmark & \checkmark & \checkmark & \checkmark & \checkmark &    \\ \cline{2-14}
     
     & Simulator of interventions for COVID-19 \cite{Simulator-of-interventions-for-COVID-19} & & & & & \checkmark &  & \checkmark & \checkmark & \checkmark &  & &  \\ \cline{2-14}
     
      & Agent-based transmission model of COVID-19 \cite{re-opening_policy_design_ABM} & & &  & & \checkmark & \checkmark & \checkmark & \checkmark & & & & \\ \cline{2-14}

      & AbCSim \cite{City-scale_model_for_COVID-19} & & & \checkmark & \checkmark &  &  & \checkmark & \checkmark & \checkmark &  & \checkmark & \checkmark  \\ \cline{2-14}
     
     & \textbf{AVSim - Airborne model (Our Approach)} & \checkmark & \checkmark & \checkmark & \checkmark & \checkmark & \checkmark & \checkmark & \checkmark & \checkmark & \checkmark & \checkmark &  \\ \hline
    \end{tabular}
    }
\end{table*}

To overcome limitations in existing models, we present AVSim, an agent-based framework for simulating both airborne and vector-borne disease dynamics under realistic conditions. Unlike approaches based on synthetic mobility, AVSim derives agent behavior from real-world GPS data by identifying stay regions and abstracting them into reusable location types. It captures both spatial and temporal aspects of mobility, where individuals go and how long they stay, stratified by profession and context. Behavioral subtypes within occupations are uncovered via unsupervised spectral clustering, enabling nuanced agent modeling. These insights are embedded in a scalable simulation environment with hierarchical geography and integrated public transport. AVSim supports detailed disease modeling through an extended SEIR (Susceptible-Exposed-Infectious-Recovered) framework and is adaptable to a range of transmission modes. A preliminary version of this work appeared in \cite{our1, our2}; this article expands on it with improved methodologies, additional experiments, and validation through simulations of COVID-19 and dengue propagation.

In summary, we make the following contributions to the agent-based modeling community:
\begin{itemize}
  \item A novel human motion modeling approach that leverages GPS data and density-based spatial clustering to generate high-resolution daily agent trajectories.
  \item Uncover heterogeneous behavioral patterns within the same occupation groups using unsupervised clustering and graph-theoretic analysis.
  \item Introduction of an agent-based model with realistic, scalable environments and transportation systems.
  \item A case study demonstrating the capability of the proposed ABM in simulating the spread of COVID-19 and dengue.
\end{itemize}

The rest of this paper is organized as follows:
Section \ref{sec: related work} reviews related work on disease modeling and human motion modeling using ABMs.
Section \ref{sec: method} details the methods used to model and simulate human behavior within ABM.
Section \ref{sec: results} presents the results, discussion, and validation of our approach.
Finally, Section \ref{sec: conclusion} concludes by highlighting the significance of the proposed framework.

\section{Related Work}
\label{sec: related work}

Numerous methods have been employed to model disease spread and uncover patterns to better understand disease propagation so that effective disease control strategies can be adopted. These models can be broadly classified into two categories: compartment models and agent-based models (ABMs). Compartment models are relatively simple and computationally efficient. Within the compartmental modeling framework, the SEIR model and its modified versions have been widely used to study epidemics and pandemics \cite{gallagher2017comparing, SMC-Lightweight-Social-Computing, SMC-A-Binary-Particle-Swarm-Optimizer-for-Networked-Epidemic-Control}. However, they fail to present granular, micro-level details about the simulation.
ABMs, on the other hand, are more complex and able to deliver micro-level details about each agent in the simulation. They have the capability to model interactions between agents or entities in a defined environment. Each agent follows a set of well-defined rules, allowing ABMs to capture emergent behaviors that arise from these interactions. This makes them particularly useful for modeling complex systems in fields like epidemiology, healthcare systems, disaster response, manufacturing and automation, transportation and mobility, economic and market behavior ~\cite{COVASIM, OPENABM, 7874209, 6202710, 9788029, 6202709}. The ability to incorporate fine-grained details enhances decision-making, especially for policymakers seeking accurate predictions. Some benchmark ABMs are mentioned and compared with our proposed ABM in Table~\ref{tab:disease_simulator_comparison}.

Additionally, there has been increasing research on the dynamic analysis of vector-borne diseases. One notable approach by Carrie Manore et al.~\cite{patch-2014} employs an ABM to analyze the propagation of mosquito-borne diseases. This study introduces the network-patch method, utilizing differential equations to model mosquito density. Similarly, Imran Mahmood et al.~\cite{dengueMohomad} use an ABM to compute vector density based on the reproductive behavior of vectors. However, their model does not incorporate realistic movement patterns, limiting its ability to provide real-world insights. Another study applied a fuzzy modeling approach to develop an epidemic model for mosquito-borne diseases \cite{dengueDAYAN2022105673}. Many of these models have modeled agents by representing them as nodes in a network ~\cite{COVASIM, OPENABM}. In these networks, human behavior and interactions were modeled using probability distributions~\cite{SMC-An-ACP-Approach-to-Public-Health-Emergency-Management}. Although reasonably accurate, these methods fail to capture the full extent of human motion dynamics. 

\section{Material and methods}
\label{sec: method}
\begin{figure*}[h]
    \centering
    \includegraphics[width=1\linewidth]{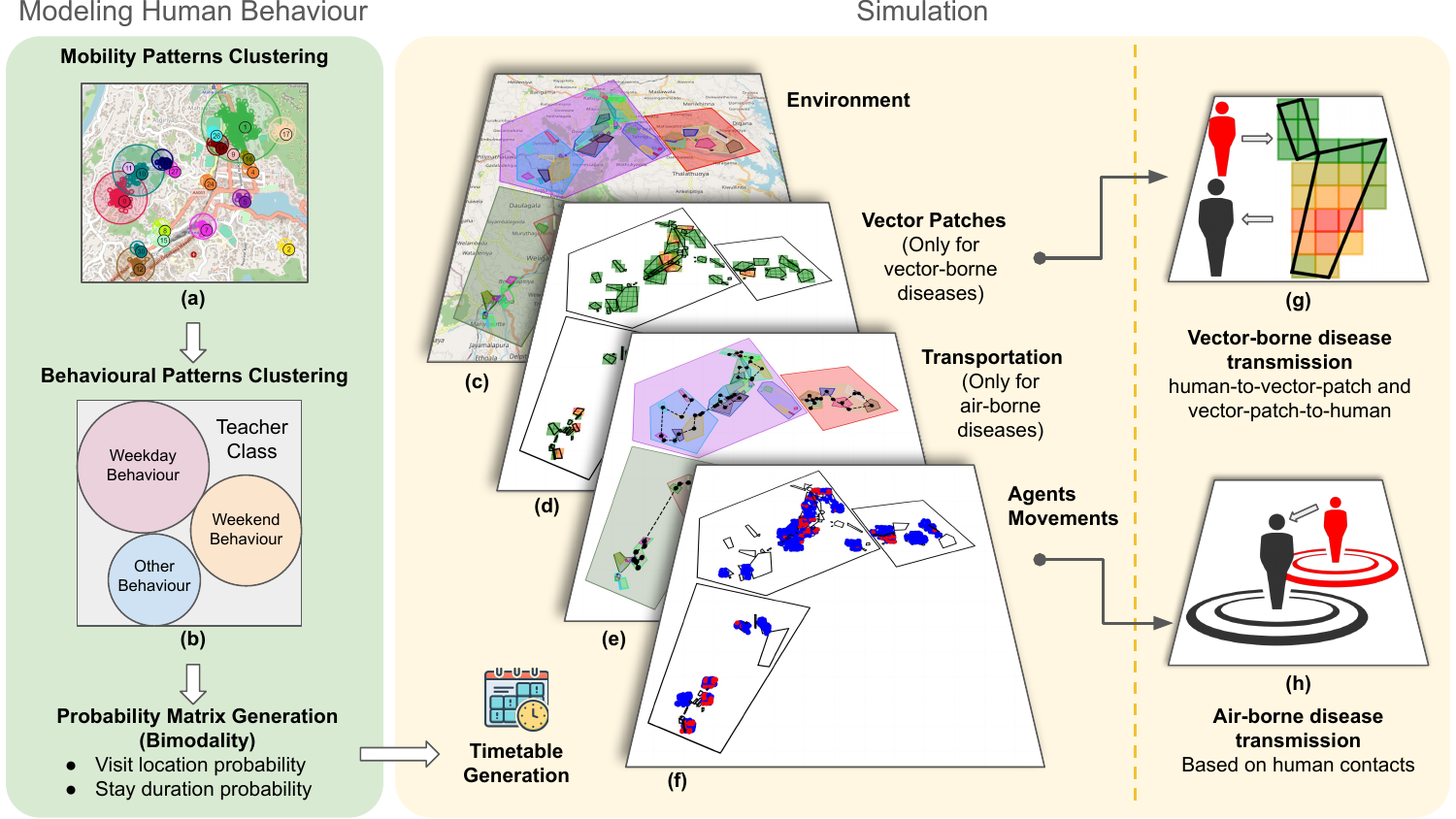}
    \caption{Overview of the proposed agent-based model structure, illustrating the key components: environment, vector-patch, transportation, agent simulation, and the two disease transmission models: vector-borne and airborne.}
    \label{fig: methodology}
\end{figure*}

The proposed design of AVSim consisted of two major stages. The first stage was the data processing stage, where human mobility and behavioral patterns were identified and generated using real-world data. This sets the foundation for the second stage of AVSim: the design of the agent-based modeling system. As shown in Fig.~\ref{fig: methodology}, the ABM is comprised of four main components: the environment, vector patches, transportation network, and agents. Each of these components will be explained in detail in the following sections.

\subsection{Modeling human behaviour}
\label{method:mobility patterns}
To accurately model disease spread in a human population, it was essential to replicate realistic human motion patterns within the simulation environment. We identified the following factors that affect human motion: 

\begin{itemize}
  \item Individuals visit specific locations based on the time of day.
\item The length of time spent at a location varies depending on the time of day and the nature of the location.
\item The types of locations visited and the associated durations differ based on
individuals’ professions.
\end{itemize}

To incorporate these factors, we introduced a novel GPS-based approach using real-world mobility data from 100 volunteers in Kandy, Sri Lanka, representing 13 professions. These profession classes were selected such that they capture the broader population of Kandy District. Location coordinates, altitude, and timestamps were recorded at 5-minute intervals over 21 days. Therefore, the dataset consisted of a total of 2100 days. Since the ABM simulator (described in subsequent sections) operates at a 1-minute resolution, the data was upsampled using Zero-Order Hold reconstruction. 

{\bf Ethical approval}: This study was approved by the Ethical Review Committee, Faculty of Arts, University of Peradeniya, Sri Lanka (ARTS/ERC/2021/01, 18 September 2021), with support from the Department of Sociology. Administrative clearance was granted by Sri Lanka's Ministry of Home Affairs, relevant Divisional Secretariat offices, and village officer divisions.

\subsubsection{Mobility pattern clustering}

To generalize the geographically specific GPS data collected in Kandy, Sri Lanka, we transformed the coordinate-based (longitude/latitude) data into a time-location domain representation. This abstraction preserves essential mobility patterns while removing dependence on physical coordinates.

\begin{figure}[h]
\includegraphics[width=1\linewidth]{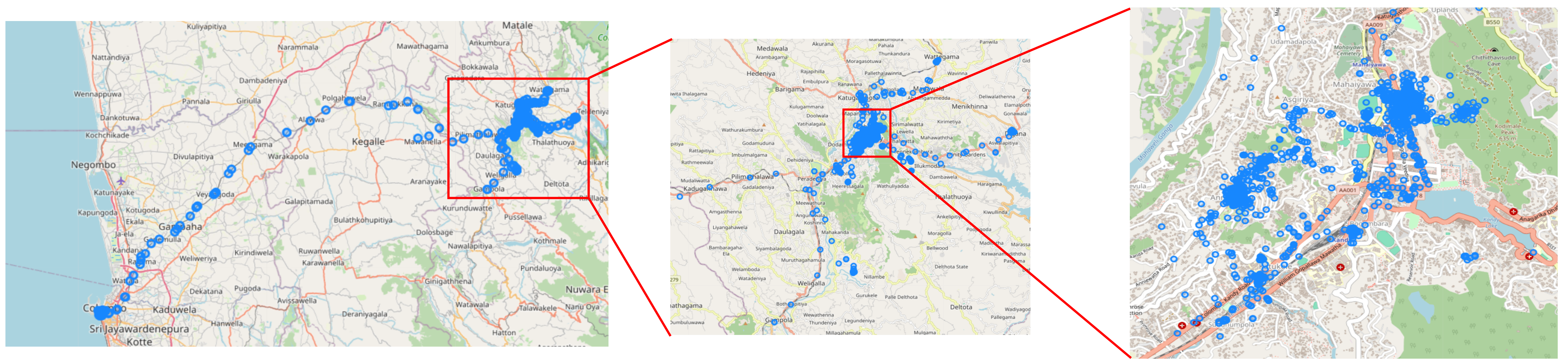}
    \caption{Hierarchical clustering of locations as we zoom into GPS data.}
    \label{clusterZoom}
\end{figure}

To identify the locations visited by each person, the GPS data for all participants over the 21 days was plotted on a map of Sri Lanka. When zooming in on these denser clusters (which occur at locations where the person pauses), additional sub-clusters emerge within them, revealing further granularity. Fig.~\ref{clusterZoom} illustrates this inherent pattern in human motion, highlighting a hierarchical structure in the locations visited, with primary locations further divided into more specific destinations. 

This study focuses on denser clusters of GPS data, indicating locations where participants spent more time. We used the Density-Based Spatial Clustering of Applications with Noise (DBSCAN) algorithm to detect these clusters, as it handles varying densities and is robust to noise \cite{DBSCAN-1, DBSCAN-2}.

DBSCAN requires two parameters: epsilon (\(\epsilon\)), the maximum distance between points (in meters), and minPts, the minimum number of points to form a cluster. We set (\(\epsilon\)) = 5 m and minPts = 10, which produced the most consistent and meaningful clusters.

To map clusters to real-world locations, we labeled them using predefined locations and zones. For example, if a cluster represented an individual's home, it was labeled as "Home." Similarly, clusters corresponding to areas around the individual's neighborhood were labeled as "Residential Zone."

\begin{figure}[h]
    \includegraphics[width=1\linewidth]{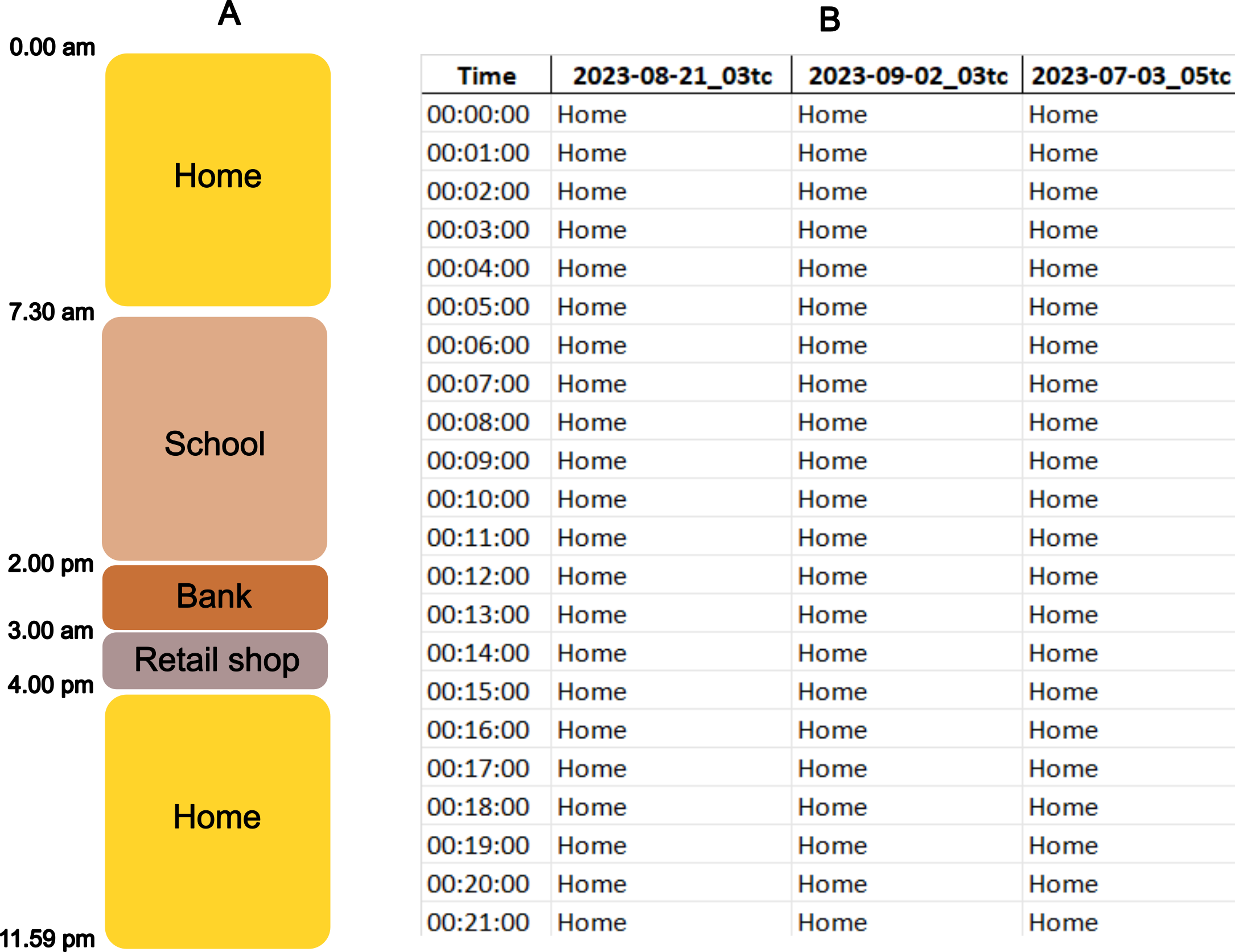}
    \caption{Final dataset: A visual representation of the trajectory of a teacher for a regular workday (A) and a part of the actual final dataset (B).}
    \label{Traj}
\end{figure}

 Transportation points were excluded from clustering and do not appear in the final dataset. A part of the reconstructed dataset with a visual summary is shown in Fig.~\ref{Traj}. The statistics needed to model human motion were derived from this dataset.

\subsubsection{Behavioral patterns clustering}
\label{spectral clustering}
The processed dataset revealed occupation-based societal structures, with distinct behavioral sub-classes emerging within each profession, as shown in Fig.~\ref{fig: methodology} (b). We applied Spectral Clustering \cite{chung1997spectral} to identify these hidden patterns, building on methods from \cite{ng2002spectral,Ranasinghe2022}.

For effective clustering, we numerically encoded locations using a 6-digit binary system where the first three digits uniquely identified zones and the remaining three digits specified locations within those zones. This encoding ensured a clear separation between distinct locations, while locations within the same zone had less separation, reflecting their relative proximity or similarity.

Let us consider a day as a data point. Then we can represent our data set $(X)$ as, $X = x_{1}, x_{2},...,x_{k}, \dots, x_{n}$ where $x_{k}$ is the GPS data of $k\textsuperscript{th}$ day and $n$ is the total number of days. Consider the notion of similarity between $x_{i}$ and $x_{j}$ as $S_{ij} > 0$.

To define similarity, we have chosen a Gaussian kernel, as shown below. The Gaussian kernel assigns high similarity to close-by points, allowing them to cluster together, while maintaining clear separation between distinct groups.
\begin{equation}
 S_{i,j} =\begin{cases}{exp(-\frac{\parallel{x_{i}-x_{j}}\parallel^{2}}{2\sigma^2})} & i \neq j\\\mu & i = j\end{cases}   
\end{equation}
Here, sigma ($\sigma$) is a scaling parameter that determines the closeness of data points to be considered a cluster. The other parameter, $\mu$, is a predefined value. In this study, it will be referred to as self-similarity, which will be in the range of [0, 1]. In our case, we selected $\mu$ as 1 since it yielded the best possible results. 

\begin{figure}[h]
    \centering
    \includegraphics[width=1\linewidth]{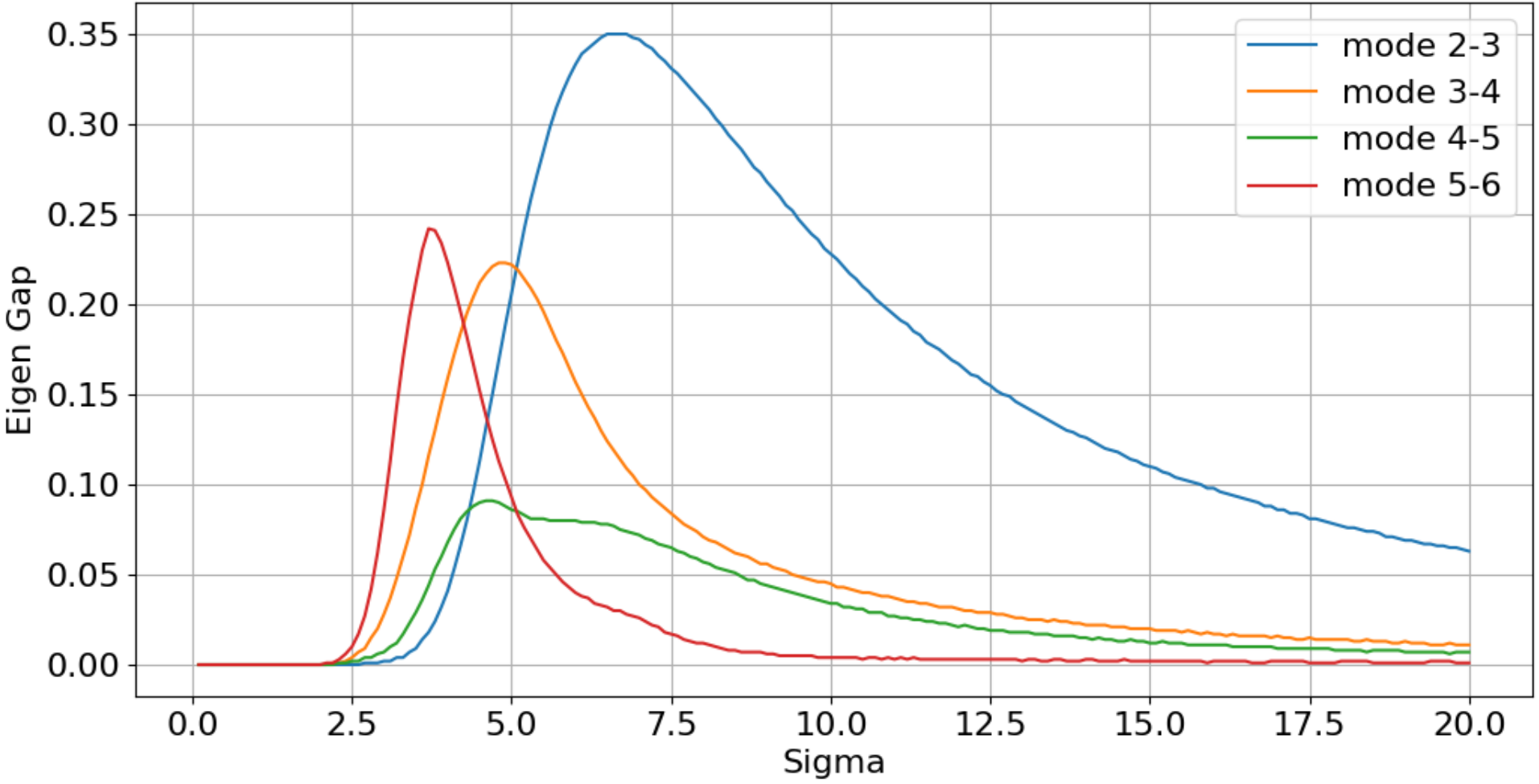}
    \caption{Sigma sweep for bank worker occupation class.}
    \label{sigma_sweep}
\end{figure}

First, a similarity matrix (Adjacency matrix) was formulated as shown below.
\begin{equation}
A_{n,m} = 
\begin{pmatrix}
S_{1,1} & S_{1,2} & \cdots & S_{1,n} \\
S_{2,1} & S_{2,2} & \cdots & S_{2,n} \\
\vdots  & \vdots  & \ddots & \vdots  \\
S_{m,1} & S_{m,2} & \cdots & S_{m,n} 
\end{pmatrix}
\end{equation}
\noindent Next, we created the degree matrix as,
\begin{equation}
D_{[i,j]} =
\begin{cases}{\sum\limits_jA_{[i,j]}} & i = j\\\mu & otherwise.
\end{cases}
\end{equation}
\noindent Then, the symmetric normalized Laplacian was obtained,
\begin{equation}
L=I-D^{-1/2}AD^{-1/2}.
\end{equation}
Following eigen decomposition of $L$, we sorted eigenvalues in ascending order and analyzed eigen gaps versus $\sigma$ as shown in the sigma sweep plot (Fig.~\ref{sigma_sweep}). The mode $x-y$ corresponds to the eigen gap between $x\textsuperscript{th}$ and $y\textsuperscript{th}$ eigenvalues after sorting.

Using the aforementioned sigma sweep plot as a reference, the number of optimal clusters can be determined by applying the selection algorithm outlined below:

\begin{enumerate}
  \item  The eigen gap with the highest value was considered as the prominent mode.
  \item Selection of Dominant Sigma: Once the prominent mode is determined, the $\sigma$ value corresponding to the highest gap value for this mode is selected as the dominant $\sigma$ according to the selection algorithm \cite{Ranasinghe2022}.
  \item Clustering Procedure: With the prominent mode and dominant $\sigma$ identified, clustering can be executed by following the steps.
\end{enumerate}

For example, when considering the sigma sweep for the bank worker class, as shown in Fig.~\ref{sigma_sweep}, the prominent mode could be determined as 2-3, which would mean the behavior would be broken down into two significant clusters at the corresponding $\sigma$ value. Then the matrix, 
\begin{equation}
V=[v_{1}, v_{2}, v_{3}, ... ,v_{n}] \in 	\mathbb{R}^{n\times{n}},
\end{equation}
can be formulated by stacking eigenvectors in columns. By treating each row of $V$ as a point, the resulting points can be clustered into the desired number of groups using K-means clustering.

The above method was proven to be very successful in identifying hidden motion patterns within each profession class. These results were used to model human behavior, which will be discussed in the next section.

\subsubsection{Modeling agent trajectories}
\label{method:bimodality}
To model human motion trajectories, we employed a robust probabilistic approach capable of capturing complex movement dynamics. This was introduced as a hypothetical model for modeling human motion by Weligampola et al.~\cite{Weligampola2023}. We enhanced and validated it using real-world GPS data, allowing us to generate daily agent trajectories based on day type (weekday or weekend) and profession within a defined environment.

\begin{figure}[h]
    \includegraphics[width=1\linewidth]{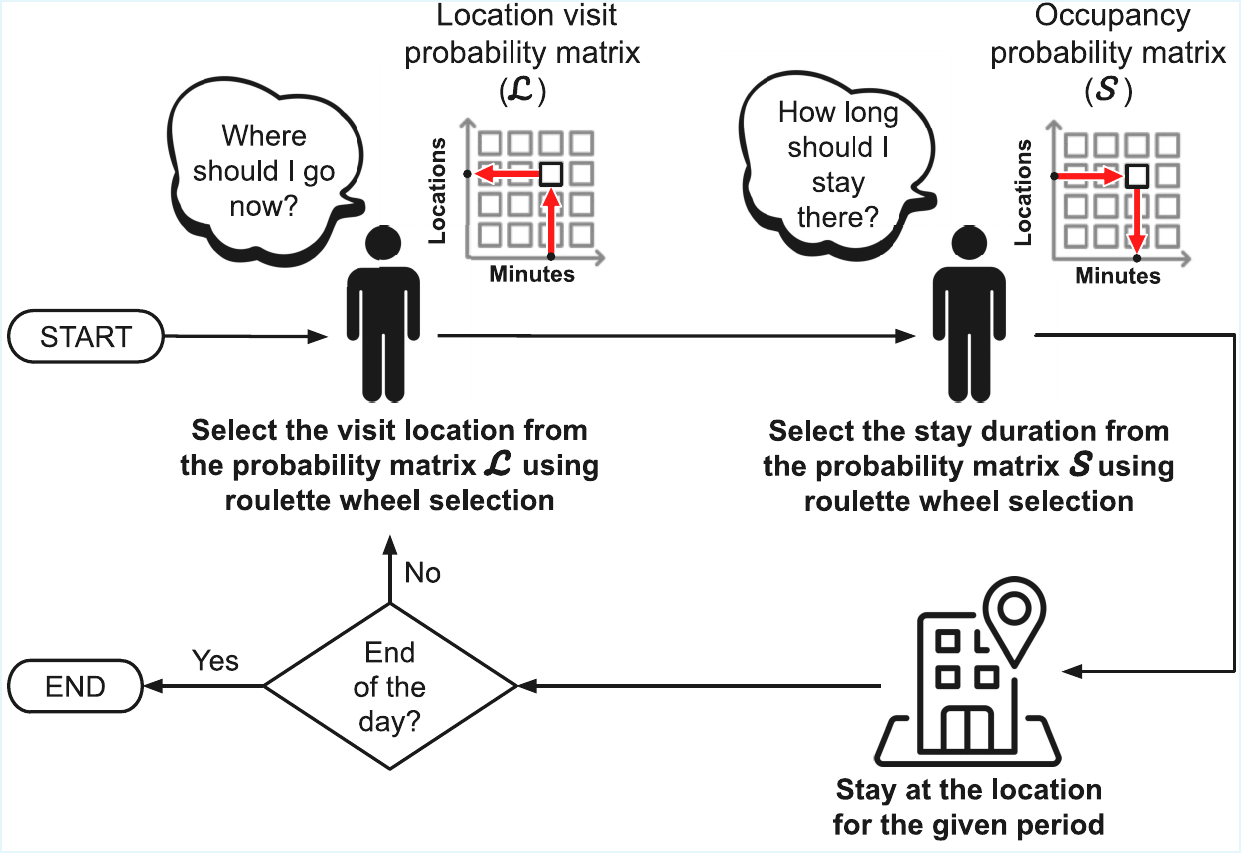}
    \caption{Process of generating daily trajectories.}
    \label{tt_process}
\end{figure}

This method was inspired by patterns observed in GPS trajectories, where denser point clusters indicated longer stays. It was also noted that human movement involves pseudo-random visits to locations with varying stay times. These observations form the basis for the theory proposed by Weligampola et al.~\cite{Weligampola2023}.

We start by introducing two probability matrices with size $n\times m$, where $n$ represents the number of locations and $m$ represents the total number of minutes in a day:
\begin{enumerate}
  \item Location visit probability matrix $(\mathcal{L})$ - Represents the probability of being at a specific location at any given minute of the day. 
  \item Location occupancy probability matrix $(\mathcal{S})$ - Represents the probability of remaining at a location for a specified duration. 
\end{enumerate}

We extracted these two matrices for each profession class using the time-location dataset generated earlier. As shown in Fig.~\ref{tt_process}, we generate trajectories through roulette wheel selection on these matrices. 

\subsection{Environment}
Given that the motion data was sourced from the Kandy District in Sri Lanka, the same environment was recreated within the ABM framework. Major cities such as Kandy, Pallekele, and Gampola were modeled using polygons on a real-world Map and further divided into sub-cities and functional zones (e.g., educational, medical, residential). For example, nearby schools were grouped into one educational zone, as shown in Fig.~\ref{fig: methodology} (c).

\begin{figure}[h]
    \centering
    \includegraphics[width=1\linewidth]{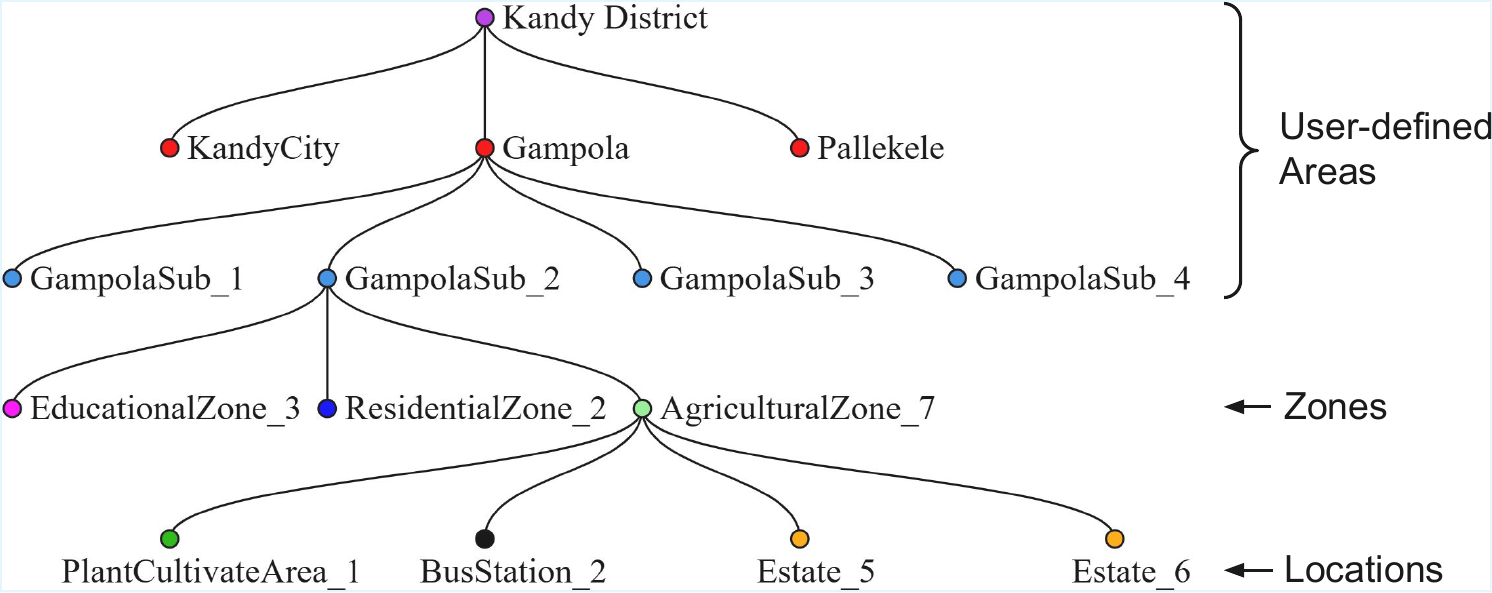}
    \caption{Hierarchical tree structure of the environment.}
    \label{fig:tree}
\end{figure}

To represent this geographic hierarchy, we used a tree-based structure in the ABM, enabling agents to move realistically between locations. Each node in the tree represents an area and is implemented as an object with attributes defining its characteristics (Fig.~\ref{fig:tree}, showing one expanded zone for clarity).

\subsection{Agent generation and simulation}

In AVSim, agents were first grouped by occupation and then further divided based on behavioral patterns. For example, day-shift and night-shift doctors were treated as separate groups. Each agent was then assigned parameters that defined their characteristics, including occupation class, sub-class (if applicable), age ranges, home and work location classes, and other factors necessary for disease modeling. This agent composition was selected to approximately represent their real-life counterparts. Thereafter, the simulation process was initiated. Simulation steps were executed every 1 minute for each day (resolution of the simulation). Each day began with timetable generation for agents (see Section \ref{method:bimodality}), after which they followed their schedules using available transport modes.

\subsection{Transport} 

In the context of disease modeling, long-range human mobility plays a critical role in facilitating the rapid geographical dissemination of emergent infectious diseases \cite{brockmann2009human}. Our study incorporates the simulation of long-range human mobility by modeling both private and public transportation systems to better understand their impact on disease spread.   

Whenever an agent had to travel from one location to another, the traveling path in the simulation environment(intermediate cities and zones the agent passed by) was obtained using the graph object shown in Fig ~\ref{fig:tree}. For this, the "traveling salesman problem" algorithm ~\cite{TSP} was used.

Public transport modeling holds significant importance in the context of disease modeling, particularly due to the increased vulnerability of passengers to airborne disease transmission in confined spaces with close proximity to others \cite{CUMMINGS2024111303}. 

\begin{figure}[h]
    \includegraphics[width=1\linewidth]{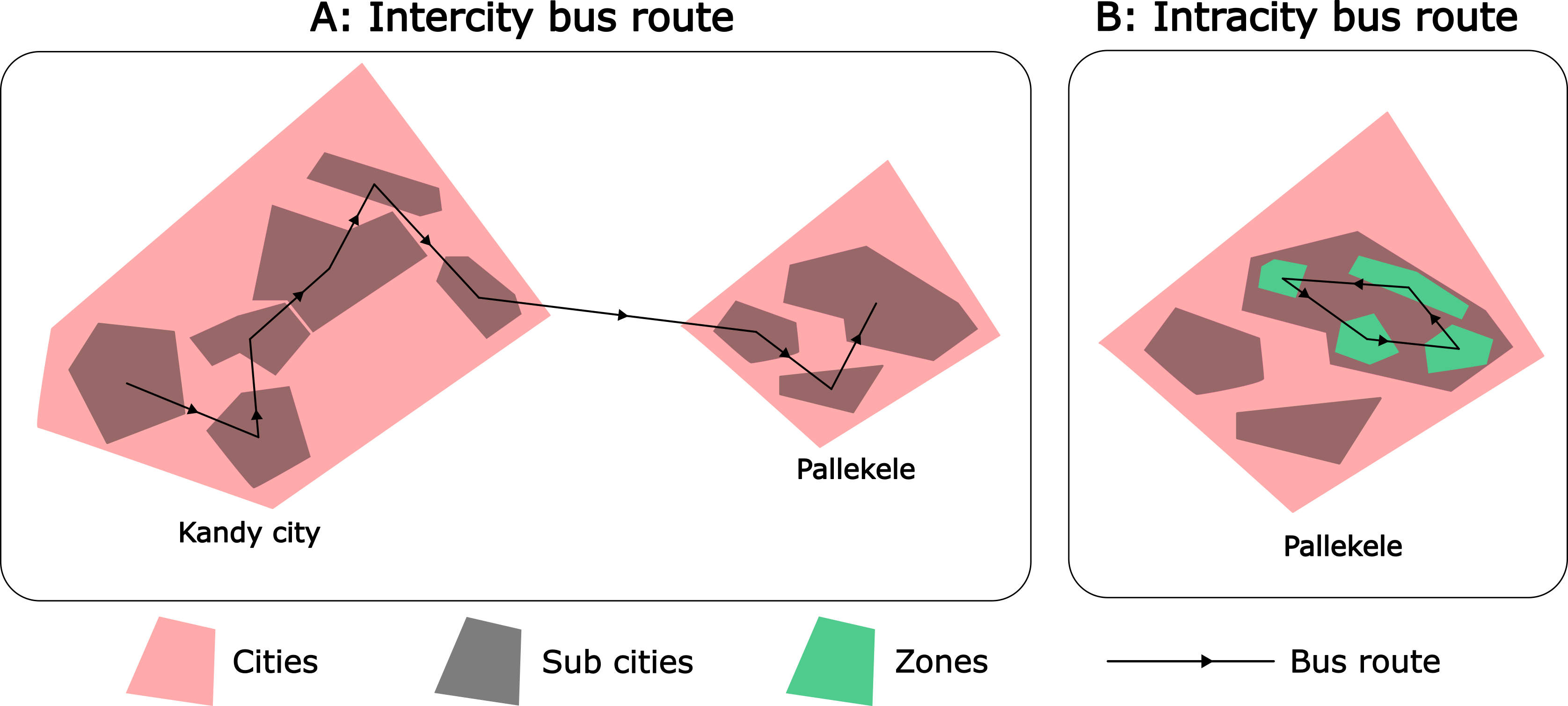}
    \caption{Public transport mechanism.}
    \label{transportr}
\end{figure}

In this study, buses and taxis were used as the primary modes of public transport. Two types of bus objects were employed to represent different hierarchical levels in the simulated environment. Inter-city buses facilitate transport between cities, while intra-city buses facilitate transport within a city. This process is explained in Fig.~\ref{transportr}.

\subsection{Vector patch}
\label{method:patch}

In agent-based models of vector-borne diseases, it's unnecessary to model each vector individually. Instead, representing vector density within localized areas offers a simpler yet effective approach. Aedes aegypti mosquitoes, for example, usually stay near human habitats and travel up to 400 meters to lay eggs \cite{whoDengueSevere}. Based on similar studies \cite{patch-2014}, we use a patch-based method to model vector density.

Each zone is divided into patches, sized according to typical mosquito habitat ranges as shown in Fig.~\ref{fig: methodology} (d).

Each patch is characterized by environmental factors such as temperature, humidity, and rainfall, which influence vector emergence and death rates. The total birth rate of vectors in a given patch $k$, $h_v^k$, is calculated as follows:
\begin{equation}
   h_v^k = N_v^k \left( \psi_v - (\psi_v - \mu_v) \cdot \frac{N_v^k}{K_v} \right) 
\end{equation}
Here, $\psi_v$ is the average natural emergence rate, $\mu_v$ is the average death rate, $K_v$ is the carrying capacity, and $N_v^k$ is the total number of vectors in patch $k$.

\subsection{Markov disease progression model}
\label{method:markov}
\begin{figure}[h] 
\centering \includegraphics[width=1\linewidth]{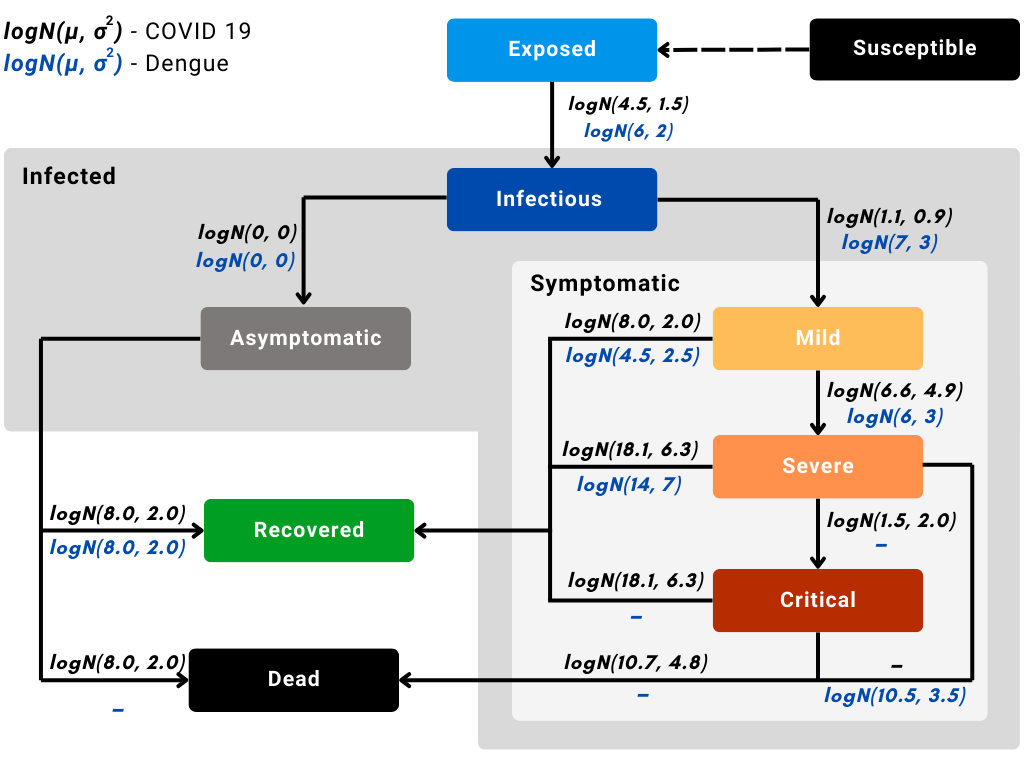}
\caption{Markov model for disease transmission modeling.} 
\label{fig:markov model} 
\end{figure}

After exposure to a disease, a person transitions through different states based on factors such as disease type, age, prior infections, and medical care. Typically, an exposed individual becomes infectious, potentially spreading the disease asymptomatically or with symptoms ranging from mild to critical. We can model this using a Markov model (Fig. \ref{fig:markov model}), with transition times following a log-normal distribution \cite{Weligampola2023}. Data for COVID-19 were taken from Covasim~\cite{COVASIM} while dengue was taken from WHO~\cite{whoDengueSevere}. 

\subsection{Airborne disease modeling}
\label{air borne method}

\subsubsection{Disease progression mechanism}

Initially, a small percentage of the population was infected with the COVID-19 virus, simulating natural disease spread as they moved freely. Transmission occurred either at specific locations or on public transport. At each time step, a 1-meter radius around each agent was monitored. If another agent enters this radius, it is recorded as a contact. Disease transmission between agents is governed by their transmission probabilities, \(\rho\), which is influenced by the agent's immunity. In the simulation, agents were assigned varying immunity levels based on their age, hygiene practices (such as social distancing, hand washing, and sanitizer use), and vaccination status. The infection probability is modeled as:
\begin{eqnarray}
\label{eq:infec_prob_AirB}
\rho = S_{age} \cdot k \cdot (1 - \alpha_{vacc} \gamma_{vacc} - \alpha_{hyg} \gamma_{hyg})
\end{eqnarray}
where \(S_{age}\) represents age-based susceptibility, \(\alpha_{vacc}\) and \(\alpha_{hyg}\) are the respective weights for vaccination and hygiene, and \(\gamma_{vacc}\) and \(\gamma_{hyg}\) represent the immunity boosts provided by vaccination and hygiene practices. \(k\) is a tunable parameter. 

Values for \(S_{age}\)  were taken from Covasim \cite{COVASIM}.
Through some prior tuning processes, the value of \(k\) was selected as 0.3. 

\subsubsection{PCR testing and quarantine mechanism}

Polymerase Chain Reaction (PCR) testing is crucial in predicting the risk of airborne diseases \cite{West2008-xo}. In AVSim, this could be used as a pre-event testing activity, and people with positive PCR results were identified as infected agents. In the COVID-19 pandemic, infected agents were identified using PCR tests, and exposed agents were identified by tracing the paths of contacts of infected agents. Both infected and exposed agents were then subjected to quarantine or self-isolation. Therefore, AVSim could be used by policymakers to study the effect of various PCR and quarantine policies.

\subsubsection{Vaccination mechanism}

Vaccination mechanisms are utilized to enhance community immunity, particularly during outbreaks of airborne diseases. In AVSim, vaccination events could be organized to reduce the disease spread while increasing the immunity levels of the agents. The AVSim simulation allowed the organization of vaccination events on a selected day at a specified geographical zone. AVSim further allowed the capability to vaccinate targeted occupational classes and give out various types of vaccines. When an agent is given a vaccination, their vaccine immunity \(\gamma_{vacc}\) is boosted. 

\subsection{Vector-borne disease modeling}
For modeling vector-borne diseases, the network-patch approach proposed by Carrie Manorea et al.~\cite{patch-2014} for dengue transmission has been adapted in this study to reflect the dynamics more accurately. The transport model was excluded, as the likelihood of contracting a vector-borne disease during transit is minimal. Nevertheless, the agents’ motion dynamics were retained.

\subsubsection{Disease progression in vectors}
\label{method:vb-disease-propagation}
As mentioned in Section \ref{method:patch}, we do not consider individual vectors separately. Instead, we used a patch to represent vector density and employed a stochastic process to model disease progression. 
Specifically, we focused on three sequential disease state transitions of the vectors: susceptible, exposed, and infected.

Each patch $k$ is initialized with counts for susceptible ($S_v^k$), exposed ($E_v^k$), and infected ($I_v^k$) vectors, ensuring that these values remain below a defined vector carrying capacity specific to each patch. The temporal dynamics of vector density within a patch $k$ are modeled using a system of differential equations as follows.
\begin{equation}
  \frac{dS_v^k}{dt} = h_v^k - \lambda_v^k S_v^k - \mu_v^k S_v^k  
\end{equation}
\begin{equation}
    \frac{dE_v^k}{dt} = \lambda_v^k S_v^k - v_v^k E_v^k - \mu_v^k E_v^k
\end{equation}
\begin{equation}
    \frac{dI_v^k}{dt} = v_v^k E_v^k - \mu_v^k I_v^k
\end{equation}
Here, \(h_v^k\) is the total vector birth rate, \(\lambda_v^k\) is the average infection rate, \(v_v^k\) is the rate of progression from exposed to infectious, and \(\mu_v\) is the average vector death rate. The $\lambda_v^k$, is calculated as follows:
\begin{equation}
  \lambda_v^k = b_v^k \cdot \beta_{vh} \cdot \left( \frac{I_h^k}{N_h^k} \right)  
\end{equation}
where \(b_v^k\) is the number of bites per vector per unit time, \(\beta_{vh}\) is the probability of transmission from an infectious agent to a susceptible vector upon contact, \(I_h^k\) is the number of infected agents in patch \(k\), and \(N_h^k\) is the total number of agents in that patch. The \(b_v^k\), is calculated as:
\begin{equation}
    b_v^k = \frac{b^k}{N_v^k}
\end{equation}
where \(b^k\) is the total number of vector-human contacts (bites) per unit time in patch \(k\).
\begin{equation}
    b^k = \frac{\sigma_v \cdot N_v^k \cdot \sigma_h \cdot N_h^k}{\sigma_v \cdot N_v^k + \sigma_h \cdot N_h^k}
\end{equation}
Here, \(\sigma_v\) is the maximum number of bites per vector per unit time, \(\sigma_h\) is the number of bites a human can receive per unit time, \(N_v^k\) is the number of vectors, and \(N_h^k\) is the number of agents in patch \(k\).

For the selected use case of dengue virus infection, the average rate per agent of progression from the exposed state to the infectious state of vectors in patch $k$, denoted as $v_v^k$, depends on the temperature~\cite{Xiao2014} and was calculated as follows:
\begin{equation} 
    v_v^k = \frac{1}{tinc_v^k} 
\end{equation}
where \(tinc_v^k\) is the temperature-dependent incubation time of the virus in mosquitoes within patch \(k\). Based on the patch temperature \(T^k\), \(tinc_v^k\) is drawn from a uniform distribution: \(U(10, 25)\) if \(T^k < 21\), \(U(7, 10)\) if \(21 \leq T^k < 26\), and \(U(4, 7)\) if \(26 \leq T^k < 31\).

The parameters used for dengue transmission were taken from similar studies \cite{patch-2014} and the WHO website \cite{whoDengueSevere}. Since the model includes multiple vector patches, a range of values was selected for some parameters to add variability.

\subsubsection{Disease progression in agents}
The infection rate of agents, \(\lambda_{k,h}(t)\) in patch \(k\) depends on the number of bites per unit time \(b^k_h\), the proportion of infectious mosquitoes, \({I_k^v}/{N_k^v}\), and the probability of successful transmission, \(\beta_{hv}\). This relationship is given by:
\begin{equation}
    \lambda_{k,h}(t) =  b^k_h \cdot \beta_{hv} \left( \frac{I_k^v}{N_k^v} \right)
\end{equation}

Probability of an exposed agent becoming infected within a given time step is calculated as follows. The simulation is executed in time steps of $\Delta t$ (1 min in our case), while the disease states of agents and vectors are updated every $n \cdot\Delta t$ (5 min) time steps. Assuming the time to infection follows an exponential distribution \cite{patch-2014}, the probability of infection at the end of the time interval $\Delta t$ is given by:

 \begin{equation}  
p_{k} = 1 - e^{-\lambda_{k,h} \Delta t}
 \end{equation}

Once an agent is exposed, the disease transitions follow the Markov model as shown in Fig.~\ref{fig:markov model}. An infected person may either stay asymptomatic, develop mild symptoms, or, in rare cases, progress to severe symptoms \cite{whoDengueSevere}. The likelihood of transitioning between these states depends on various factors, including age, prior dengue infections, and the quality of medical care. For instance, the likelihood of an infected person developing symptoms may vary based on their immune response.

The Markov model outlined in Section \ref{method:markov} is used to determine state transitions. For dengue, the model focuses only on mild and severe symptomatic states. By combining probabilistic state transitions with random transition periods, the model can effectively simulate diverse disease progression outcomes among individuals.

\subsubsection{Hospitalization mechanism}
There is a high probability of symptomatic patients visiting the hospital. At the start of each day in the simulation, all agents' disease statuses are checked. If symptomatic agents are present, they will be hospitalized based on some hospitalization probability. For severe cases, this probability is higher. 

Once the agents requiring hospitalization are identified, they are isolated from other agents and put into hospitals. It is assumed that hospitalized agents do not contribute to further disease spread. However, their disease states continue to update as outlined in Fig. \ref{fig:markov model}. Eventually, these agents either recover or die. Recovered hospitalized agents are returned to both the environment and their respective patches.

\subsubsection{Intervention mechanism}
When there is a rise in reported vector-borne cases in a specific area, vector control interventions are initiated, often spearheaded by government healthcare programs or community efforts. These measures typically include mosquito spraying or the removal of breeding sites. 

In AVSim, to simulate such interventions, the total mosquito-to-host exposure in each patch is recorded. If the cumulative number of exposed agents in a zone over one week surpasses a predefined threshold, $E_h^k$, the vector population in the affected zone's patches is reduced by $m\%$ to represent mosquito spraying and the elimination of breeding sites. 

\section{Results and discussion}
\label{sec: results}

\subsection{Spectral clustering results}
\label{result:spectral clustering}
As discussed in ~\ref{spectral clustering}, distinct behavioral patterns within each occupation class were derived by applying the Spectral Clustering algorithm to each occupation class. This was done by manual inspection of the output of the clustering algorithm. Table~\ref{Occupation_Subclusters} presents the synthesized results.
Note that normal behavior refers to a typical workday for each class.
\begin{table}[htbp]
\centering
\caption{Distinct behavioral patterns obtained for selected occupation classes through the spectral clustering method.}
\label{Occupation_Subclusters}

\begin{tabular}{|>{\raggedright\arraybackslash}p{0.12\linewidth}|>{\raggedright\arraybackslash}p{0.74\linewidth}|}
\hline
\multicolumn{1}{|c|}{Class}          & \multicolumn{1}{c|}{Identified Clusters}                                                                \\ \hline

\multirow{5}{*}{Doctor}              & Normal workday behaviour\\ \cline{2-2} 
                                     & Night to evening shift                                                                                  \\ \cline{2-2} 
                                     & Night to morning shift                                                                                  \\ \cline{2-2} 
                                     & Weekend and holiday behaviour\\ \cline{2-2} 
                                     & Night shift at hospital, daytime at school medical center \\ \hline
\multirow{2}{*}{Farmer}              & Work area is near the residence\\ \cline{2-2} 
                                     & Work is conducted within an Agricultural Zone.\\ \hline

\multirow{2}{*}{\begin{tabular}[t]{@{}l@{}}Garment\\Worker\end{tabular}}      & Work area is located separately from the residence.\\ \cline{2-2} 
                                     & The individual resides in or close to the work area.\\ \hline

\multirow{3}{*}{Student}             & Normal behaviour of a student travelling from home\\ \cline{2-2} 
                                     & Weekend and holiday behaviour with irregular school attendance\\ \cline{2-2} 
                                     & Hostel student                                                                                          \\ \hline
\multirow{4}{*}{\begin{tabular}[t]{@{}l@{}}Super\\ Market\\Worker\end{tabular}} & Morning to night shift behaviour                                                                        \\ \cline{2-2} 
                                      & Afternoon to night shift behaviour\\ \cline{2-2} 
                                     & Morning to evening shift behaviour\\ \cline{2-2} 
                                     & Works morning to afternoon and evening to night, with a break in between; includes holiday behavior \\ \hline
\multirow{3}{*}{Teacher}             & Teacher resides in or near the school premises\\ \cline{2-2} 
                                     & Normal behaviour of a teacher travelling from home                                                        \\ \cline{2-2} 
                                     & Weekend and holiday behaviour\\ \hline
\end{tabular}
\end{table}

After examining the results, it can be observed that the
algorithm excels at producing clusters with distinct behavioral
patterns, ranging from regular work routines to holiday behaviors in alignment with real-world intuition. The final results and the human motion dataset we created to fuel the AVSim can be found online at \cite{https://doi.org/10.5281/zenodo.13621863}.

\subsection{Validation of human motion modeling}
\begin{figure}[h]
    \centering
    \includegraphics[width=1\linewidth]{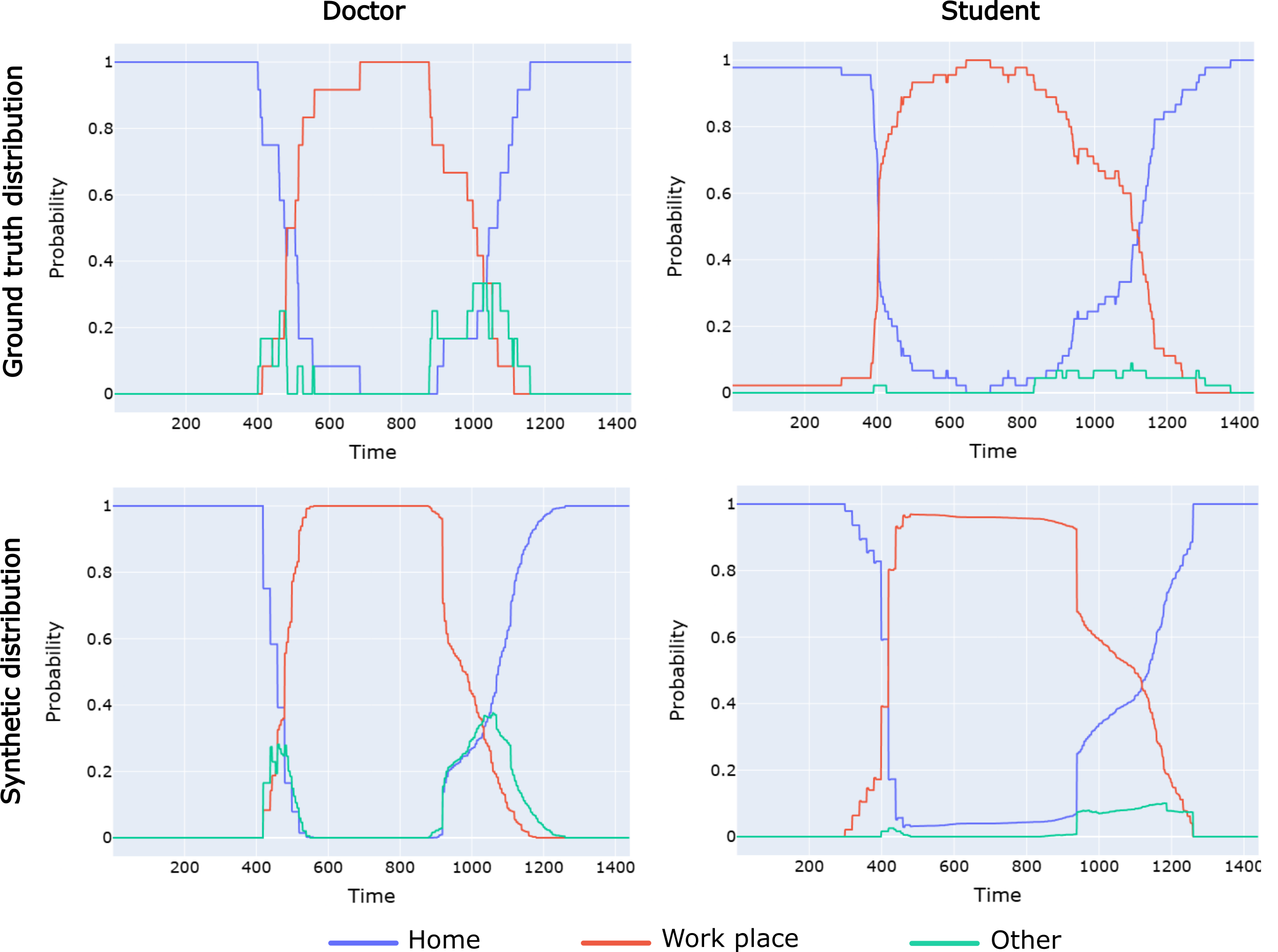}
    \caption{Validation of our motion model (Synthetic) against real world motion data (Ground truth).}
    \label{BimodelityValidation}
\end{figure}

To validate the accuracy of the human motion model, 10000 samples of daily person trajectories were generated for each profession class. From this synthetic data, the probability distributions for showing up at different locations with time were reconstructed to compare with the original ground truth probability distributions. These synthetic and original plots for the doctor and teacher classes can be seen in Fig.~\ref{BimodelityValidation}. We can see that the synthetic distributions closely resemble the ground truth distributions, thus proving the above method to be highly accurate in replicating human behavior.

\subsection{Airborne disease simulation results}
\label{air borne results}

\begin{figure*}[h]
    \centering
    \includegraphics[width=1\linewidth]{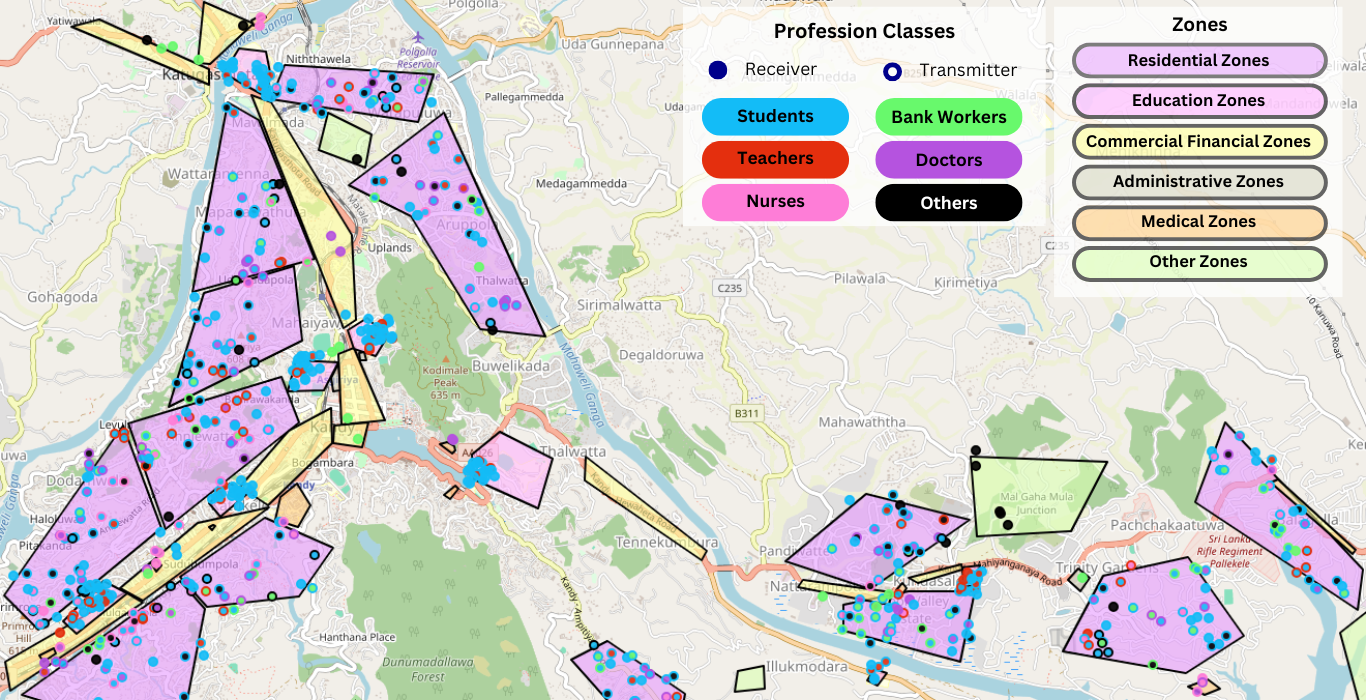}
    \caption{All the contacts in an uncontrolled simulation for 50 days in a part of "Kandy City" and "Pallekele". Initially, three students were infected randomly. The receiving agents are shown by filled circles and the transmitting agent is shown by outer borders in each instance.}
    \label{contactPlot}
\end{figure*}

In this section, we discuss the simulation results to research air-borne disease spread. However, it is important to note that these results only constitute a subset of possible research directions. One can perform various analyses based on their need and interests. All these simulations consisted of approximately 2000 agents. This is just for demonstration and, based on the magnitude of the simulation environment, the agent count could be increased at the cost of computational power.

\subsubsection{Tracing path of disease spread}

\begin{figure*} 
    \centering
  \subfloat[\label{AB_spread-a}]{%
       \includegraphics[width=0.33\linewidth]{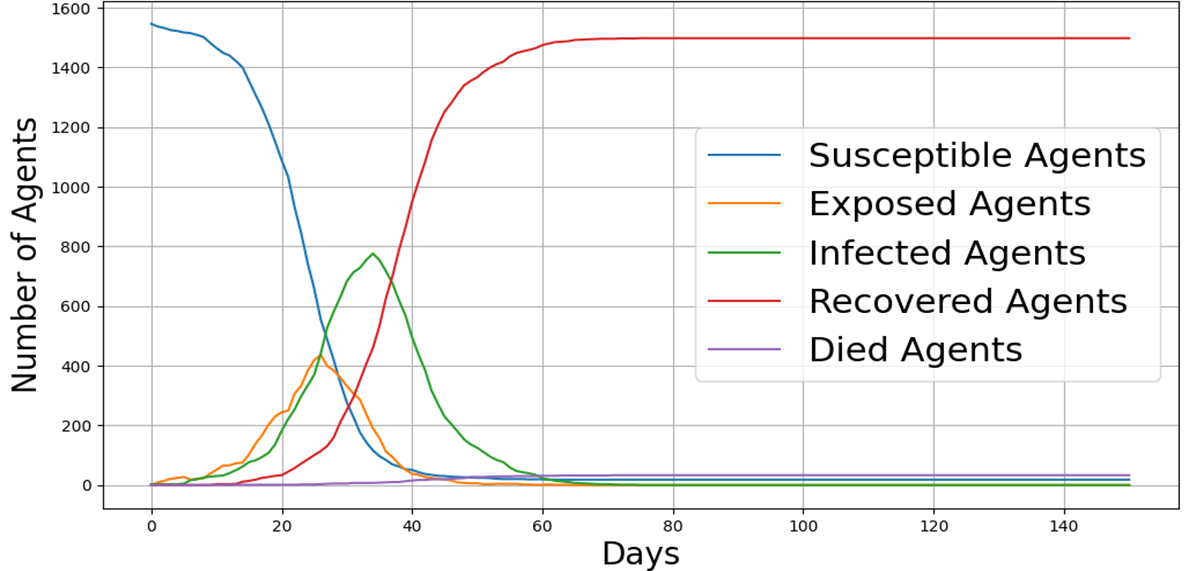}}
    \hfill
  \subfloat[\label{AB_spread-b}]{%
        \includegraphics[width=0.33\linewidth]{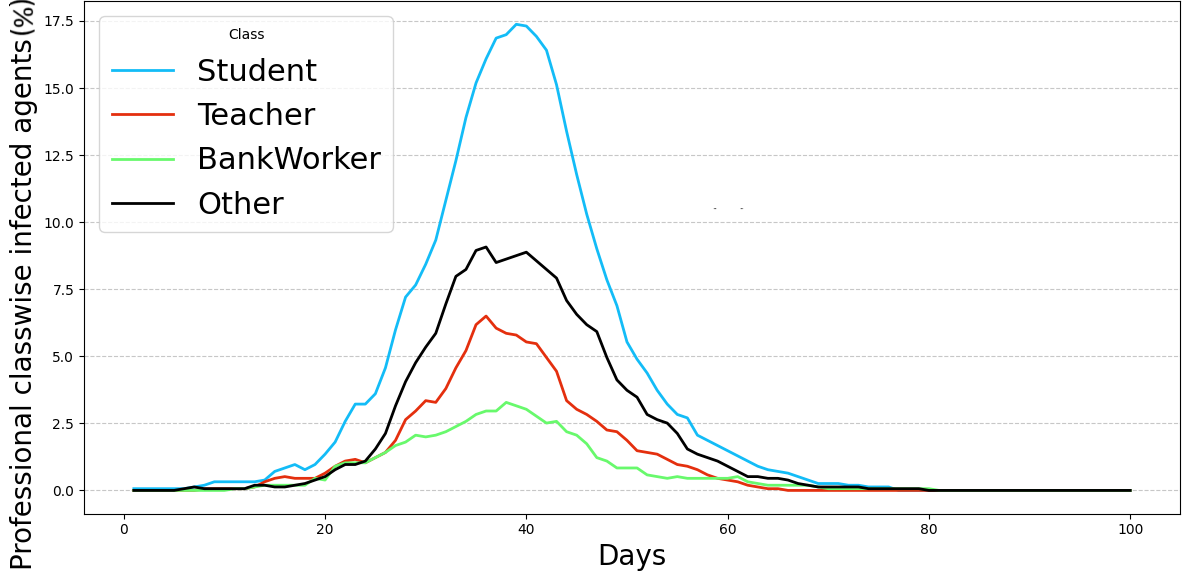}}
    \hfill
  \subfloat[\label{AB_spread-c}]{%
        \includegraphics[width=0.33\linewidth]{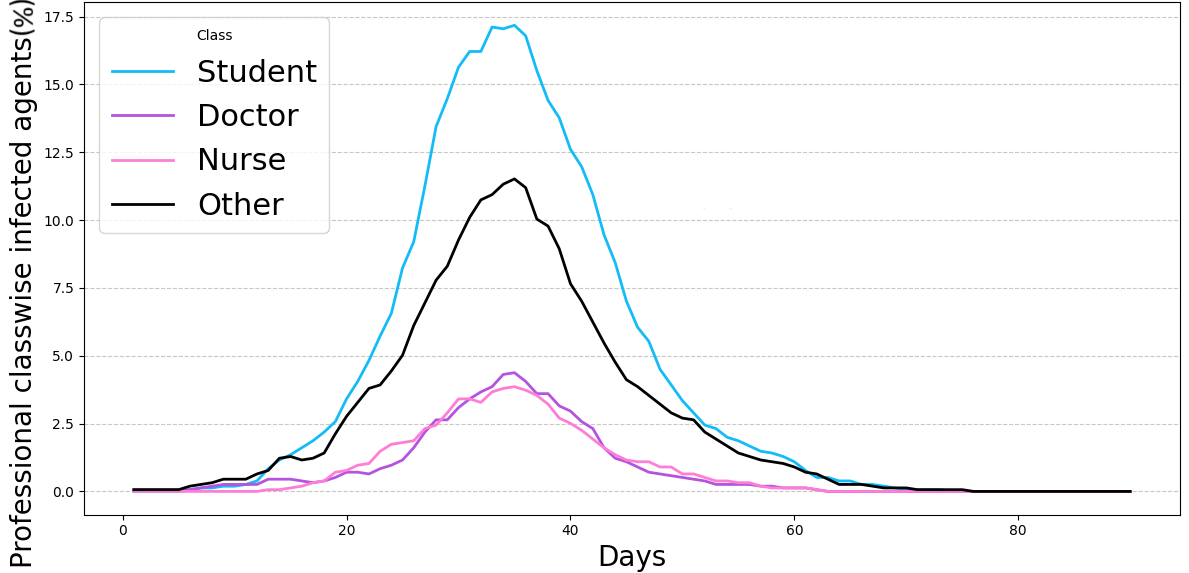}}
    \vspace{-2mm}
    \\
  \subfloat[\label{AB_spread-d}]{%
       \includegraphics[width=0.33\linewidth]{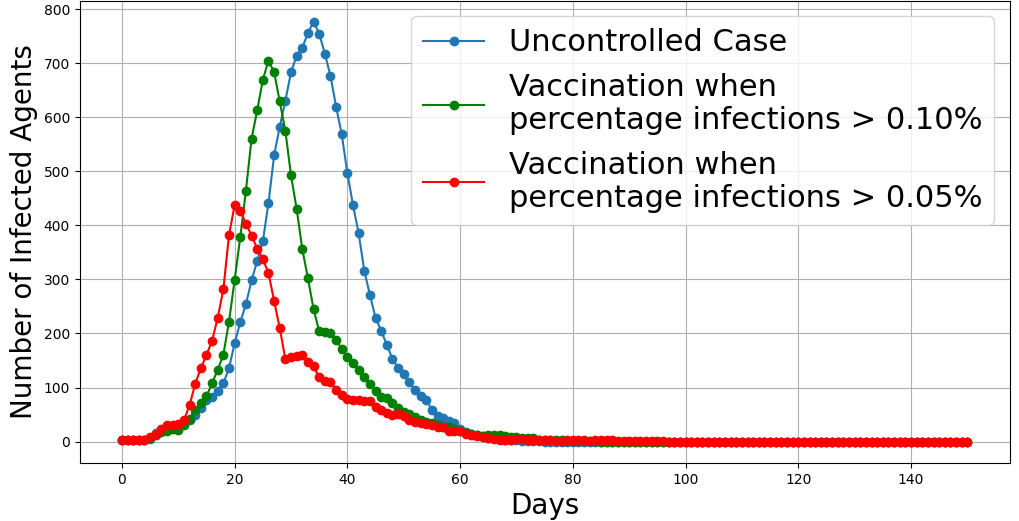}}
    \hfill
  \subfloat[\label{AB_spread-e}]{%
        \includegraphics[width=0.33\linewidth]{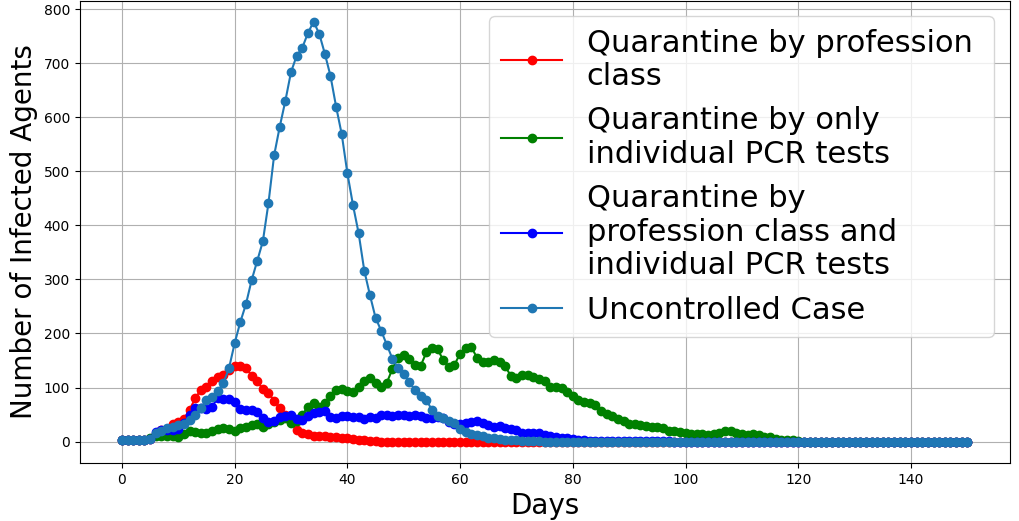}}
    \hfill
  \subfloat[\label{AB_spread-f}]{%
        \includegraphics[width=0.33\linewidth]{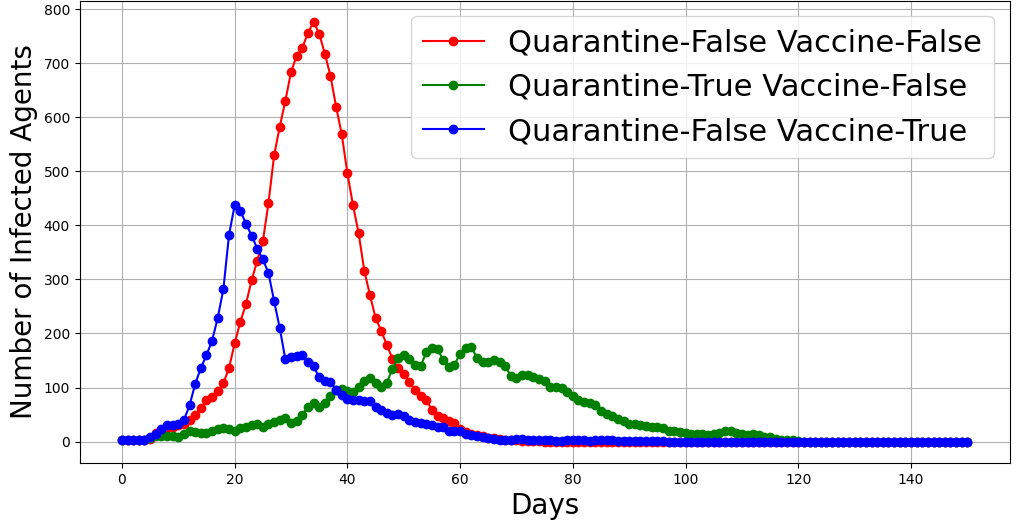}}
  \caption{(a) Agents count in different states in an uncontrolled epidemic outbreak which started with three infected students, Disease spread in different professional classes in an uncontrolled epidemic outbreak which started with (b) three infected students, (c) three random infected agents, (d) Effect of vaccine mechanism in airborne diseases w.r.t. uncontrolled case, (e) Effect of quarantine and vaccination mechanism in airborne diseases, (f) Effect of quarantine and vaccination mechanism in airborne diseases.}
   \label{AB_All_result}
\end{figure*}

\begin{figure}[h]
    \includegraphics[width=1\linewidth]{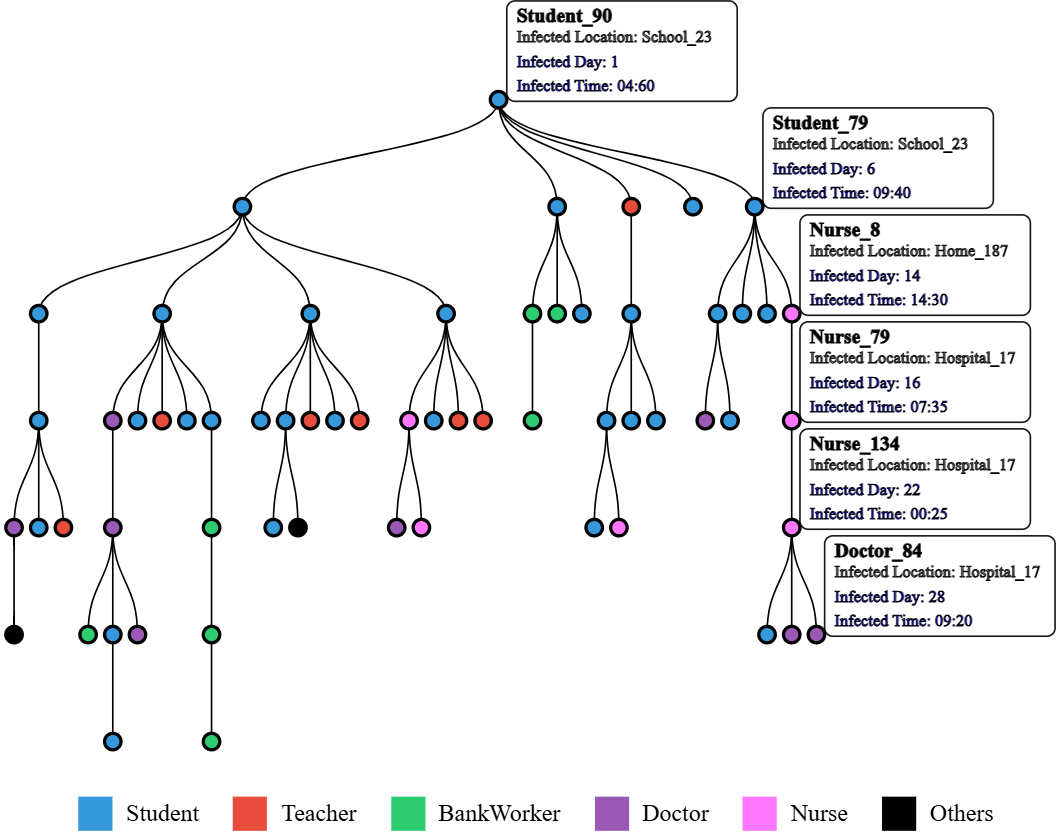}
    \caption{Tracing the paths of contacts of AVSim.}
    \label{contact_tracing}
\end{figure}

A key ABM feature is contact tracing, shown in Fig.~\ref{contactPlot}, which highlights disease transmission locations. Residential zones show more contacts, likely due to higher agent density. However, disease spread across communities often occurs through work locations. For example, in Fig~\ref{contact_tracing} an infected student transmits the disease at school, which then reaches a nurse at home, leading to further spread at the hospital where the nurse works. This illustrates how workplaces link communities in transmission chains. Leveraging AVSim, the early identification of such patterns enables policymakers to target high-risk zones and apply timely interventions.

AVSim also generates detailed descriptions of each agent, including daily routines, person trajectories, disease progression histories, and daily contacts. This is illustrated in Fig.~\ref{contact_tracing}.

AVSim simulation system can be executed in an uncontrolled environment without imposing any policies that reduce the spread. Fig.~\ref{AB_spread-a} shows such an epidemic outbreak where three random student agents were initially infected. We can see that this plot closely resembles the theoretical SEIR model curve~\cite{SEIR-valid}, hence validating the practical approach of AVSim. This can be used as the controlled experiment for the upcoming experiments.

\subsubsection{Effect of initially infected agents on disease progression}

AVSim enables analysis of disease spread from specific agent classes. In an uncontrolled setting where three students were initially infected, the disease was spread mostly to teachers within the first 25 days (Fig.~\ref {AB_spread-b}). This is due to frequent school contacts during the first couple of weeks. Over time, infections rose in other classes, revealing vulnerable groups and the effects of uncontrolled transmission.
When agents are randomly infected, it is hard to predict a pattern of spreading to other classes, and it depends on the initially infected agents. For example, in the simulation provided by Fig.~\ref{AB_spread-c}, initially infected agents were one student, one doctor, and one teacher. For this case, the disease spread had a significant impact on the medical sector agents. However, students were consistently among the most affected in all simulations, likely due to their presence in many households. 

This demonstrates that AVSim can identify which professional classes are most vulnerable based on the outbreak’s initial state, enabling the design of targeted control strategies to limit disease spread from these high-risk groups.

\subsubsection{Study of Quarantine and vaccination measures}

We have tested several controlled mechanisms to reduce the spread of airborne diseases. The following conclusions were assumed based on the parameters that we set as described in the Section \ref{air borne method}. 

\begin{enumerate}

\item Effect of Vaccine:
    Fig. \ref{AB_spread-d} shows infections under different vaccine strategies versus no control. We analyzed the effect by vaccinating all the agents when the infected percentage of the population is greater than 5\% and 10\%. It can be deduced that early vaccination leads to greater spread reduction \cite{HUANG2024805}.
\item Effect of Quarantine:
    Fig. \ref{AB_spread-e} compares infected agent counts under different quarantine mechanisms versus no control. AVSim supports two types: profession-based quarantine and self-quarantine upon positive PCR. We find that when combined, they reduce the infection peak by ~87.5
\item Vaccination and Quarantine hybrid: 
    Fig. \ref{AB_spread-f} compares infections when vaccination and quarantine were applied independently and together. Quarantine alone reduced infections more, likely due to complete movement restrictions in the simulation. However, this idealized model doesn’t reflect real-world conditions, where quarantined individuals still require some contact for basic needs, potentially leading to higher actual transmission risk.
\end{enumerate}

The results also show that vaccination helps control the spread more quickly. An ideal strategy would combine targeted vaccination, prioritizing high-risk groups like health professionals, with effective quarantine of infected agents and areas. Since AVSim is fully customizable, it can be adapted to simulate more realistic quarantine and vaccination scenarios tailored to specific geographic or contextual settings, providing policymakers with more applicable insights.
\subsection{Vector-borne disease simulation results}
\label{vec_brn_res}

Following similar logic to section ~\ref{air borne results}, all simulations were conducted with 2000 individual agents. 

Our simulation used a 500x500 square meter patch size to represent the mosquito's activity area, as Aedes aegypti mosquitoes typically fly up to 400 meters \cite{whoDengueSevere}. The other simulation parameters are set as follows: $\psi_v = 0.3$, $\mu_v = 1/14$, $\sigma_v = 0.5$, $\sigma_h = 10$, $K_v \in [100, 200]$, $\beta_{vh} = 0.33$, and $\beta_{hv} = 0.33$.

\begin{figure} 
    \centering
  \subfloat[\label{non-hospitalized}]{%
       \includegraphics[width=0.5\linewidth]{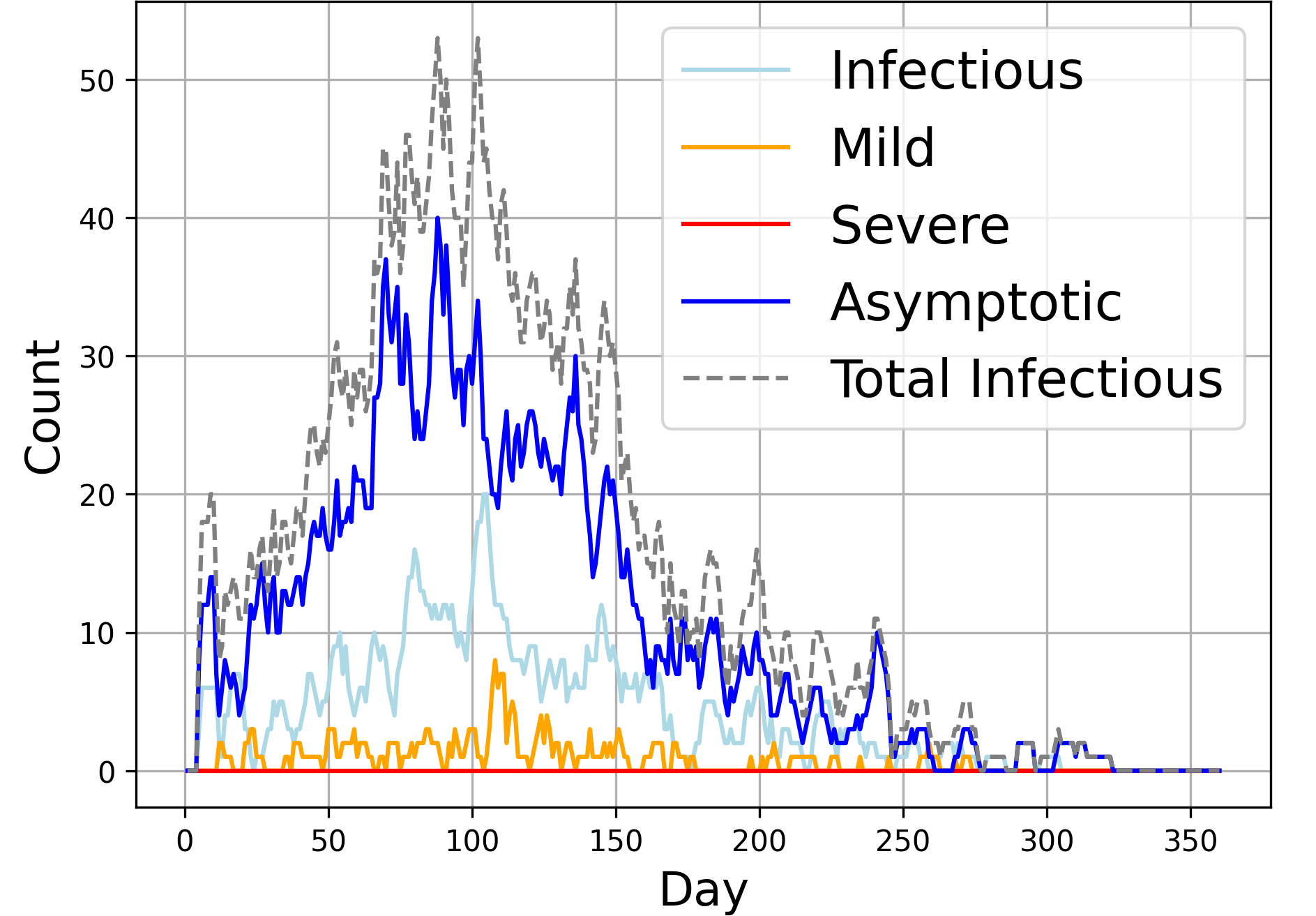}}
    \hfill
  \subfloat[\label{hospitalized}]{%
        \includegraphics[width=0.5\linewidth]{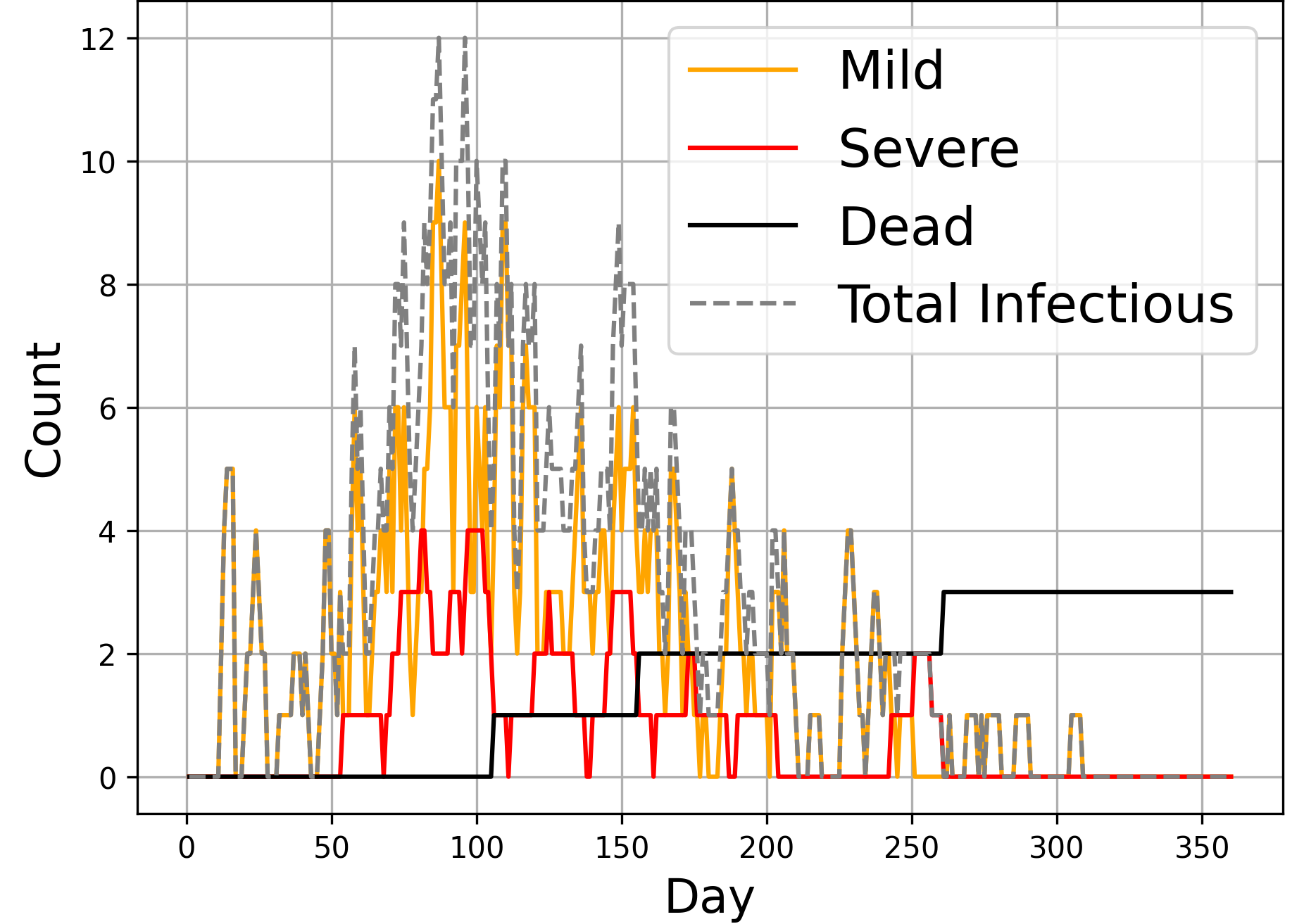}}
    \vspace{-2mm}
    \\
  \subfloat[\label{vb_disease_states}]{%
        \includegraphics[width=0.5\linewidth]{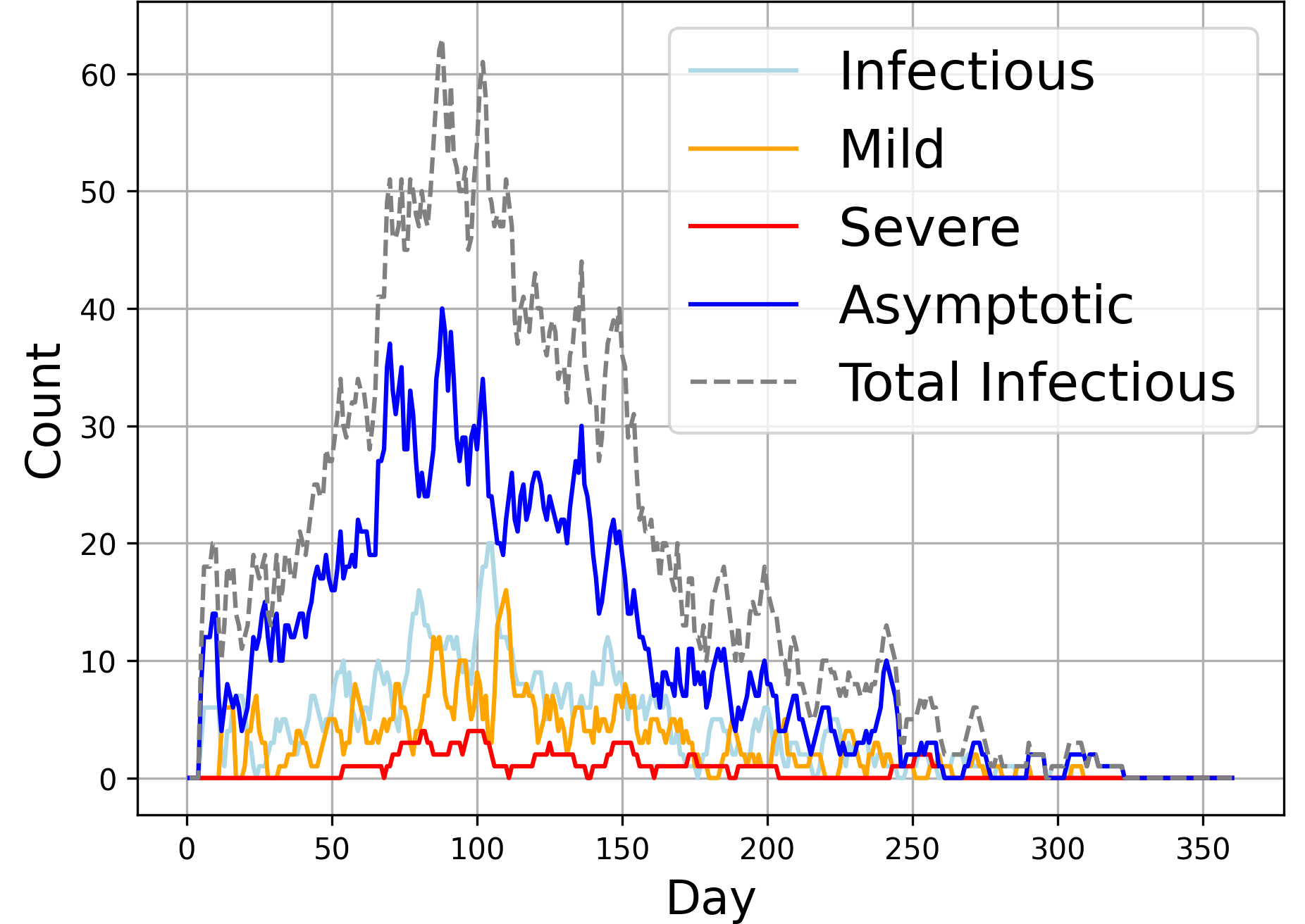}}
    \hfill
  \subfloat[\label{vb_all_states}]{%
        \includegraphics[width=0.5\linewidth]{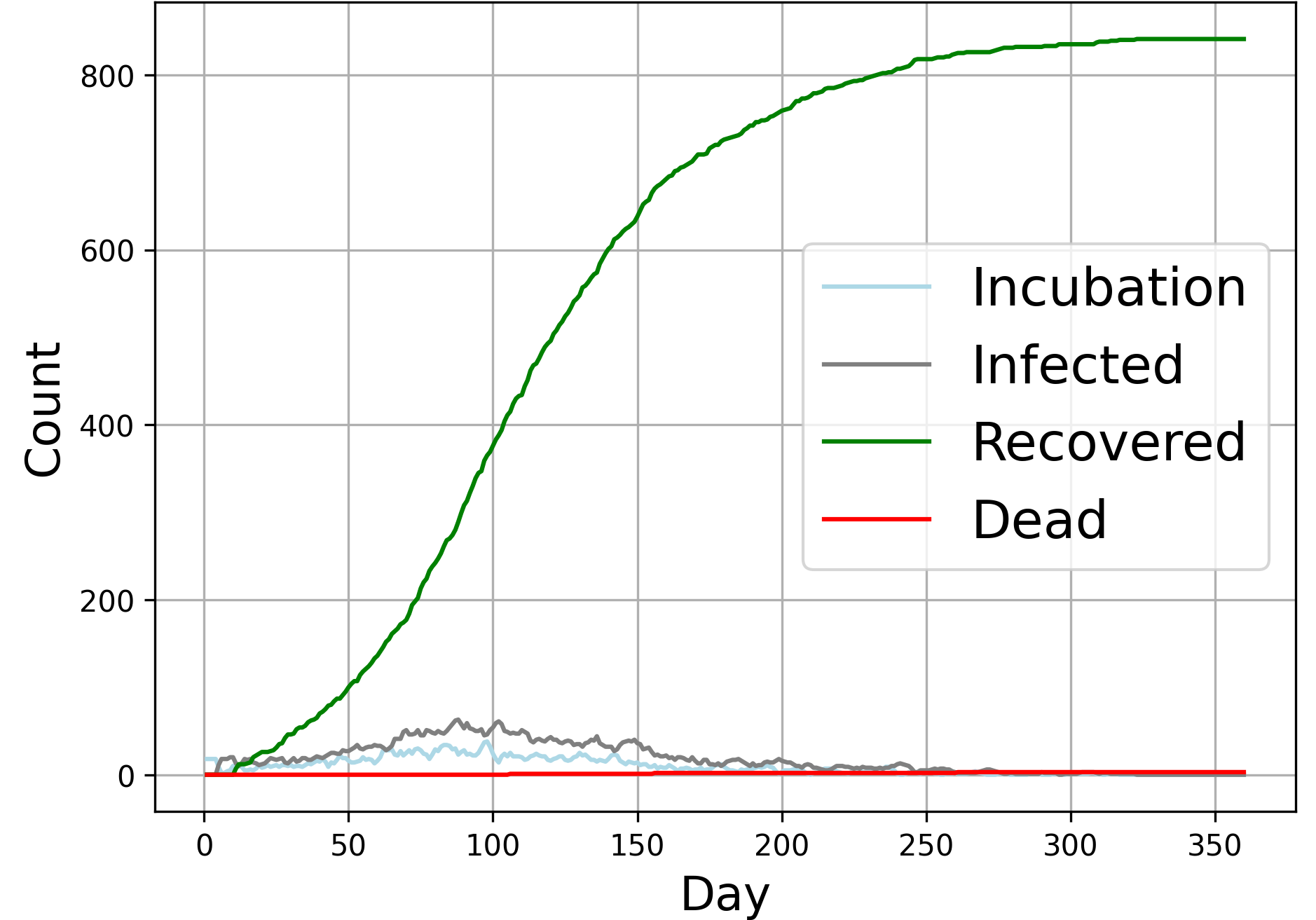}}
  \caption{Infectious disease states of (a) non-hospitalized, (b) hospitalized (only symptomatic agents are hospitalized), (c) all agents, and (d) major disease states of all agents over time.}
    \label{fig:vb-combined}
\end{figure}

\subsubsection{Disease progression in an uncontrolled environment}

At the beginning of the simulation, the patches and locations were initialized with agent and vector counts. Six homes were chosen from a designated residential zone, with up to five agents per home initially marked as exposed to dengue. The nearest nine patches around each home were given an infected vector count ranging from 0 to 20. During the simulation, the patches were updated daily, and the disease propagated accordingly.

As the simulation progresses, changes in the disease state of agents can be observed over time, as shown in Fig. \ref{fig:vb-combined}. Analyzing the rate of change indicates that once dengue starts spreading, the infection rate gradually increases before eventually declining. However, in this simulation, each individual can only be infected once. If reinfection were allowed, the infection rates would likely be higher. Reinfection could be modeled by resetting recovered agents to a susceptible state, but for simplicity, this is not considered in this simulation.

\subsubsection{Disease progression in a controlled environment}

\begin{figure}[h]
    \centering
  \subfloat[\label{controlled}]{%
       \includegraphics[width=0.48\linewidth]{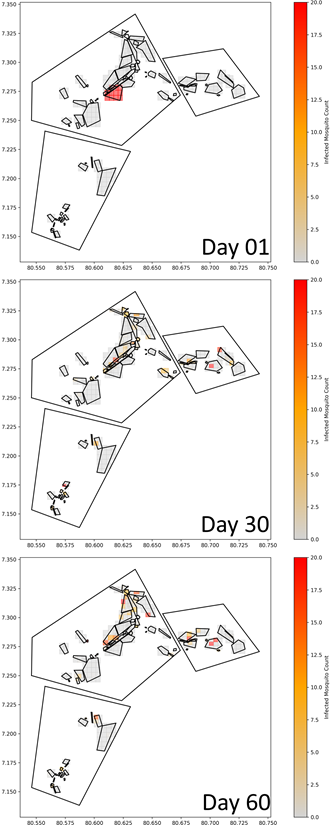}}
    \hfill
  \subfloat[\label{uncontrolled}]{%
        \includegraphics[width=0.48\linewidth]{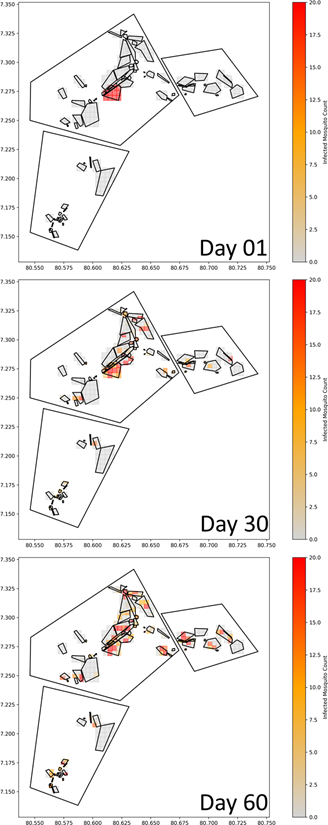}}
  \caption{The infected vector count changes over 2 months in (a) a controlled and (b) an uncontrolled environment.}
    \label{fig:mosquito-control}
\end{figure}

Various strategies can be employed to control dengue. In this simulation, the focus is on managing mosquito populations. If the cumulative number of exposed agents in a zone over an \( n \)-day period exceeds a predefined threshold, \( E_h^k \), the vector population in the affected zone's patches is reduced by \( m\% \). This reduction represents interventions such as mosquito spraying and the elimination of breeding sites. Fig. \ref{fig:mosquito-control} illustrates a comparison between uncontrolled spread and a control mechanism where \( n = 7 \), \( E_h^k = 2 \), and \( m = 75 \). 

\begin{figure}[h]
    \centering
  \subfloat[\label{fig:vb-controling}]{%
       \includegraphics[width=1\linewidth]{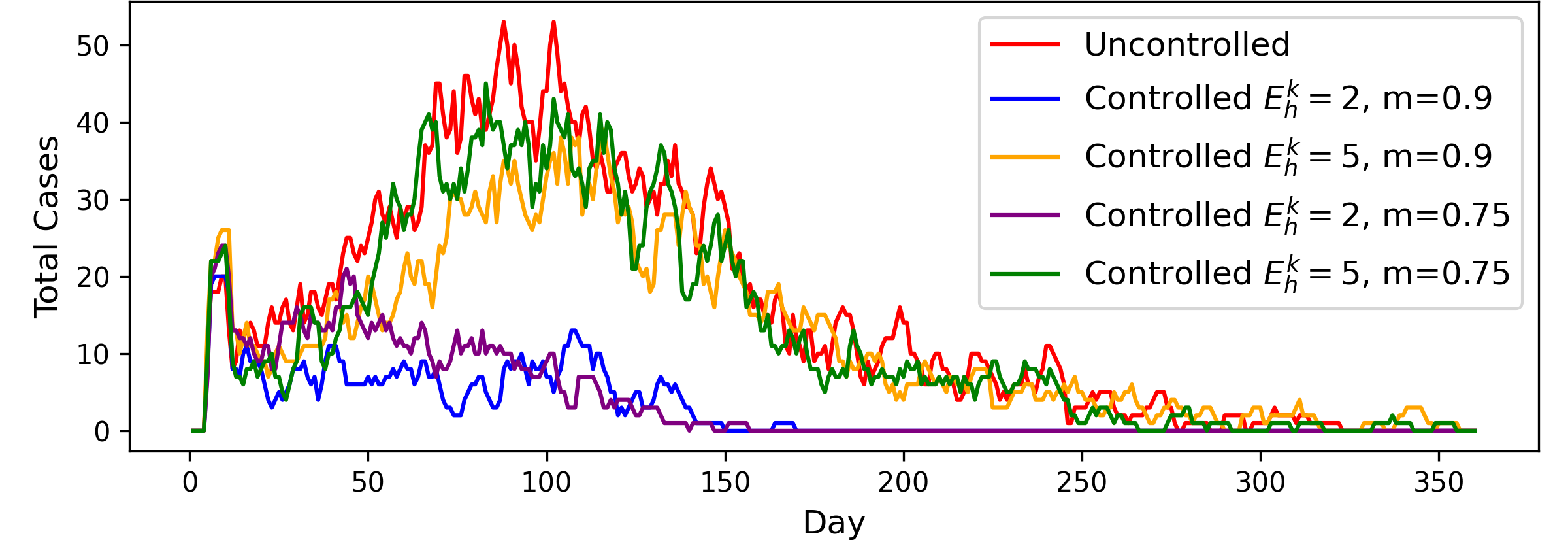}}
    \vspace{-2mm}
    \\
  \subfloat[\label{fig:vb-temp}]{%
        \includegraphics[width=1\linewidth]{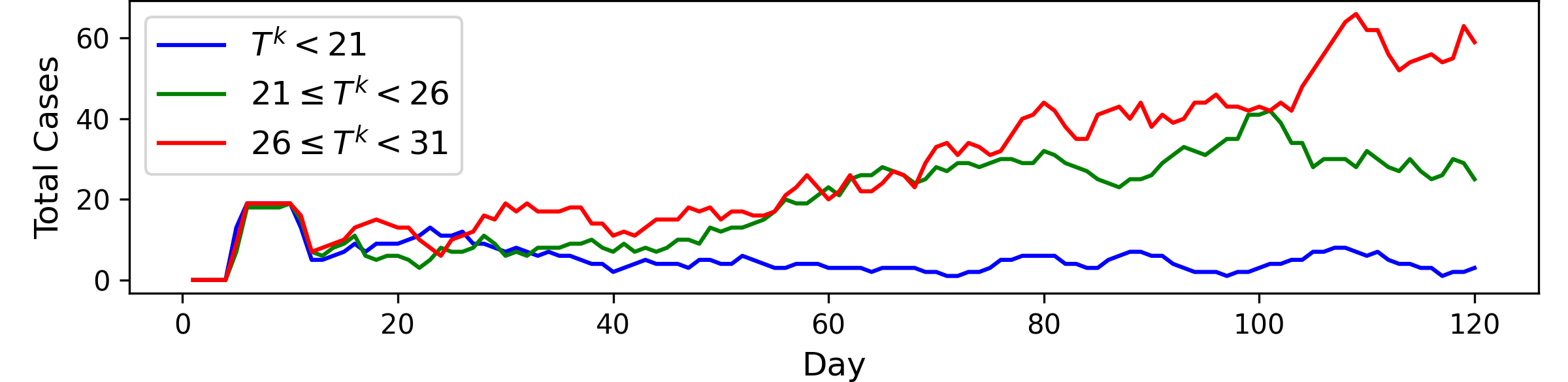}}
  \caption{The sum of Infectious, Mild, Severe, and Asymptomatic cases changes with a) the control strategies and b) temperature}
    \label{fig:temp-combined}
\end{figure}

Fig. \ref{fig:vb-controling} illustrates the mosquito control strategy under different scenarios. As the infection count increases, the number of exposed individuals in each zone also rises. At the end of each week, zones with more than n exposed agents undergo vector reduction. However, when higher thresholds for exposed counts are allowed, it becomes evident that this strategy is not significantly different from the uncontrolled case.

In contrast, when control measures are implemented even for a small number of recorded cases, the strategy proves more effective. However, this comes with an economic trade-off. By using AVSim, it is possible to determine the optimal threshold for recorded cases to trigger control measures and the extent of control needed.

In the given example, while 90\% control is the most effective, applying 75\% control can also significantly reduce mosquito populations and, consequently, dengue cases.

\subsubsection{Effect of temperature}

As mentioned in Section \ref{method:vb-disease-propagation}, the average rate of progression of vectors from the exposed state to the infectious state depends on the mosquito's incubation time. The mosquito incubation time is highly influenced by temperature. Fig. \ref{fig:vb-temp} shows how the total infectious agent count changes as the mosquito incubation period varies with temperature. During high-temperature seasons, it is important to implement necessary measures to reduce the propagation of dengue.

Similarly, environmental factors such as rainfall also play a significant role in dengue transmission. During the rainy season, increased mosquito emergence rates and higher carrying capacities contribute to a surge in dengue cases. Policymakers can analyze historical data and use simulations to predict dengue outbreaks in advance, allowing them to take necessary preventive measures for the future.

\section{Conclusion}
\label{sec: conclusion}

Airborne and vector-borne diseases, such as COVID-19 and dengue fever, pose persistent threats to humanity. Recent pandemics have exposed the severe consequences of inadequate disease control policies. Effective containment strategies require more than generalized models that rely on simulations grounded in how people move, interact, and respond across different social and spatial contexts.

AVSim addresses this need by embedding real-world mobility and behavioral patterns into an agent-based framework capable of simulating disease dynamics with occupational and environmental specificity. By grounding agent behavior in empirical GPS data and uncovering latent structure in daily routines, AVSim provides a high-resolution lens into how infections propagate through human and vector-mediated contact. Its ability to represent both airborne and vector-borne disease mechanisms while accounting for geography, transport networks, and heterogeneous behaviors enables a more nuanced assessment of transmission pathways and intervention outcomes.

Through simulations of COVID-19 and dengue, AVSim demonstrates how grounded behavioral modeling can reveal vulnerabilities in population structure, identify spatial hotspots, and evaluate the effectiveness of targeted interventions such as quarantine, vaccination, and vector control. As public health challenges become increasingly complex, AVSim provides a scalable and adaptable platform for exploring disease dynamics in a manner that is both realistic and actionable.

\bibliographystyle{IEEEtran}

\bibliography{paper-refs.bib}

\newpage
 




\vfill

\end{document}